\documentclass[a4paper,UKenglish]{lipics-v2016}
\newif\ifFull
\Fullfalse
\usepackage{t1enc}
\usepackage[utf8]{inputenc}

\usepackage{amsfonts}

\newcommand{\set}[1]{\left\{ #1\right\}}
\newcommand{\gilt}{:}
\newcommand{\sodass}{\,:\,}
\newcommand{\setGilt}[2]{\left\{ #1\sodass #2\right\}}

\newcommand{\realrange}[2]{\left[#1, #2\right]}

\newcommand{\unitrange}[2]{\realrange{0}{1}}

\newcommand{\llabel}[1]{\label{\labelprefix:#1}}
\newcommand{\labelprefix}{} 

\newcommand{\discussionsize}{\small}

\newenvironment{code}{\noindent
\begin{tabbing}%
\hspace{2em}\=\hspace{2em}\=\hspace{2em}\=\hspace{2em}\=\hspace{2em}\=%
\hspace{2em}\=\hspace{2em}\=\hspace{2em}\=\hspace{2em}\=\hspace{2em}\=%
\kill}{\end{tabbing}}

\newcommand{\labelcommand}{}
\newcommand{\captiontext}{}
\newsavebox{\codeparam}
\newcounter{lineNumber}
\newenvironment{disscodepos}[3]{%
\renewcommand{\labelcommand}{#2}%
\renewcommand{\captiontext}{#3}%
\sbox{\codeparam}{\parbox{\textwidth}{#3}}%
\begin{figure}[#1]\begin{center}\begin{code}\setcounter{lineNumber}{1}}{%
\end{code}\end{center}\caption{\llabel{\labelcommand}\captiontext}\end{figure}}

{\end{disscodepos}}

\newcommand{\Is}       {:=}

\newdimen\endofsize\endofsize=0.5em
\def\endofbeweis{~\quad\hglue\hsize minus\hsize
                 \hbox{\vrule height \endofsize width
\endofsize}\par}

\usepackage{epsfig}
\usepackage{url}
\usepackage{amsmath}
\usepackage{amssymb}
\usepackage{algorithm}
\usepackage{algorithmic}
\usepackage{pdflscape}
\usepackage{numprint}
\npdecimalsign{.} 
\usepackage{theorem}
\usepackage{makeidx}
\usepackage[usenames,dvipsnames,table,xcdraw]{xcolor}
\usepackage{hyperref}
\usepackage{xspace}
\usepackage{wrapfig}
\usepackage{graphics}
\usepackage{todonotes}
\usepackage{booktabs}
\usepackage{microtype}
\usepackage[left,pagewise] {lineno}
\definecolor {infocolor} {rgb} {0.6,0.6,0.6}

\usepackage{tikz}
\usetikzlibrary{shapes.arrows,chains}
 
\usetikzlibrary{decorations.pathreplacing}

\usepackage{framed,enumitem}

\usepackage[numbers,square,sort&compress]{natbib}

\def\MdR{\ensuremath{\mathbb{R}}}

\newcommand{\Id}[1]{\texttt{\detokenize{#1}}}

\newcommand{\ie}{i.\,e.,\xspace}
\newcommand{\eg}{e.\,g.,\xspace}
\newcommand{\etal}{et~al.\xspace}

\usepackage{color}

\newcommand{\strash}[1]{{\color{orange}[DS: #1]}}

\newcommand{\lamm}[1]{{\color{blue}[SL: #1]}}
\newcommand{\sanders}[1]{{\color{blue}[PS: #1]}}
\newcommand{\werneck}[1]{{\color{blue}[RW: #1]}}
\renewcommand{\strash}[1]{}
\renewcommand{\lamm}[1]{}
\renewcommand{\sanders}[1]{}
\renewcommand{\werneck}[1]{}

\def\comment#1{}

\def\withcomments{
  \newcounter{mycommentcounter}
   \def\comment##1{\refstepcounter{mycommentcounter}%
    \ifhmode%
     \unskip%
     {\dimen1=\baselineskip \divide\dimen1 by 2 %
       \raise\dimen1\llap{\tiny\bfseries \textcolor{red}{-\themycommentcounter-}}}\fi%
     \marginpar[{\renewcommand{\baselinestretch}{0.8}%
       \hspace*{3em}\begin{minipage}{5em}\footnotesize [\themycommentcounter]: \raggedright ##1\end{minipage}}]{\renewcommand{\baselinestretch}{0.8}%
       \begin{minipage}{5em}\footnotesize [\themycommentcounter]: \raggedright ##1\end{minipage}}}
  }

\definecolor{darkgreen}{RGB}{0,200,100}
\definecolor{orange}{RGB}{255,80,0}

\newcommand{\Xcomment}[1]{}

\newcommand{\mytitle}{ High-Quality Hierarchical Process Mapping\footnote{{This work was partially supported by the Austrian Science Fund~(FWF, project P 31763-N31) and partially supported by DFG grant FINCA (ME-3619/3-2) within
the SPP 1736 Algorithms for Big Data as well as the German Federal Ministry of
Education and Research (BMBF) project WAVE (grant 01|H15004B).}}}

\EventLongTitle{}
\EventShortTitle{}
\EventAcronym{}
\EventYear{}
\EventDate{}
\EventLocation{}
\EventLogo{}
\begin{document}
\title{\mytitle}
\author[1]{Marcelo Fonseca Faraj}
\author[2]{Alexander van der Grinten}
\author[3]{Henning Meyerhenke}
\author[4]{Jesper Larsson Träff}
\author[5]{Christian Schulz\footnote{Corresponding author.}}

\affil[1]{University of Vienna, Faculty of Computer Science, Vienna, Austria\\ \texttt{marcelo.fonseca-faraj@univie.ac.at}}
\affil[2]{Humboldt-Universität zu Berlin, Berlin, Germany \\ \texttt{avdgrinten@hu-berlin.de}}
\affil[3]{Humboldt-Universität zu Berlin, Berlin, Germany \\ \texttt{meyerhenke@hu-berlin.de}}
\affil[4]{TU Wien, Faculty of Informatics, Vienna, Austria\\ \texttt{traff@par.tuwien.ac.at}}
\affil[5]{University of Vienna, Faculty of Computer Science, Vienna, Austria\\ \texttt{christian.schulz@univie.ac.at}}

\date{}

\authorrunning{M. F. Faraj, A. van der Grinten, H. Meyerhenke, J. L. Träff, and C. Schulz}

\Copyright{M. F. Faraj, A. van der Grinten, H. Meyerhenke, J. L. Träff, and C. Schulz}
\maketitle
\begin{abstract}
Partitioning graphs into blocks of roughly equal size such that few edges run between blocks is a frequently needed operation when processing graphs on a parallel computer.  When a topology of a distributed system is known an important task is then to map the blocks of the partition onto the processors such that the overall communication cost is reduced.  We present novel multilevel algorithms that integrate graph partitioning and process mapping.  Important ingredients of our algorithm include fast label propagation, more localized local search, initial partitioning, as well as a compressed data structure to compute processor distances without storing a distance matrix.  Experiments indicate that our algorithms speed up the overall mapping process and, due to the integrated multilevel approach, also find much better solutions in practice.  For example, one configuration of our algorithm yields better solutions than the previous state-of-the-art in terms of mapping quality while being a factor~62 faster.  Compared to the currently fastest iterated multilevel mapping algorithm Scotch, we obtain~16\% better solutions while investing slightly more running time.
\end{abstract}

\section{Introduction}
\label{sec:introduction}
The performance of applications that run on high-performance computing systems depends on many factors such as the capability and topology (structure) of the underlying communication system, the required communication (patterns, frequencies, volumes, and dependencies) between processes in the given applications, and the software and algorithms used to realize the communication.  
For example, communication is typically faster (has lower latency, higher bandwidth, more communication channels) if communicating processes are located on the same physical processor node compared to cases where processes reside on different nodes.
This becomes even more pronounced for large supercomputer systems where processors are hierarchically organized (\eg islands, racks, nodes, processors, cores) with corresponding communication links of similar quality, and where differences in the process placement can have a huge impact on the communication performance
(latency, bandwidth, congestion). 
Often the communication pattern between application processes is or can be known. Additionally, a hardware topology description that reflects the capacity of the communication links is typically available.  Hence, it is natural to attempt to find a good mapping of the application processes onto the hardware processors such that pairs of processes that frequently communicate large amounts of data are~located~closely. 
Finding such best or just good mappings is the objective of some usually hard~optimization~problems.

Previous work can be grouped into two categories. 
One line of research intertwines process mapping with multilevel graph partitioning (see for example \cite{DBLP:journals/fgcs/WalshawC01,DBLP:conf/hpcn/PellegriniR96}). 
To this end, the objective of the partitioning algorithm -- most commonly the number of cut edges -- is typically replaced by an objective function that considers the processor distances. 
Throughout these algorithms, the distances are directly taken into consideration.
The second category decouples partitioning and mapping (see for example \cite{schulz2017better,brandfass2013rank,heider1972computationally,muller2013optimale}). 
First, a graph partitioning algorithm is used to partition a large graph into $k$ blocks, while minimizing some measure of communication, such as edge-cut, and at the same time balancing the load (size of the blocks). 
Afterwards, a coarser model of computation and communication is created in which the number of nodes matches the number of processing elements (PEs) in the given processor network. 
This model is then mapped to a processor network of $k$ PEs with given pair-wise distances using a process mapping algorithm. 
As shown in \cite{Traff06:mesh}, the decoupling approach can lead to worse results than the integrated approach.
We refer the reader to \cite{GPOverviewBook,SPPGPOverviewPaper} for more details on mapping and graph partitioning.

The starting point for this research is as follows. 
Recently, process mapping algorithms have made two assumptions that are typically valid for modern supercomputers and the applications that run on those: communication patterns are sparse and there is a hierarchical communication topology where links on the same level in the  hierarchy exhibit the same communication speed. 
Using these assumptions, better decoupled --~non-integrated~-- mapping algorithms have been obtained, e.g.~\cite{schulz2017better}. 
In this problem formulation, the model of computation and communication is first partitioned using a standard graph partitioning algorithm, and then a smaller model that has the same number of nodes as the underlying network of processors is mapped.
On the other hand, there has been a large body of work on the multilevel (hyper-)graph partitioning problem, which led to enhanced partitioning quality or faster local search~\cite{kaffpa,kabapeE,pcomplexnetworksviacluster,dissSchulz,schlag2016k}. 
The \emph{multilevel} approach~\cite{SPPGPOverviewPaper} is probably the most prominently used algorithm in the graph partitioning field.
Here, the input is recursively \emph{contracted} to obtain a smaller instance which should reflect the same basic structure as the input. 
After applying an \emph{initial partitioning} algorithm to the smallest instance, contraction is undone and, at each level, \emph{local search} methods are used to improve the partitioning induced by the coarser level. 
Recent enhancements to the multilevel scheme include novel local search techniques such as very localized local search algorithms, fast label propagation algorithms,  {} or gain caches to avoid expensive recomputations throughout local search algorithms.

Our \emph{main contribution} in this paper is the integration of process mapping into a multilevel scheme with high-quality local search techniques and recently developed non-integrated mapping algorithms. Additionally, we introduce faster techniques that avoid to store distance matrices.
Overall, our algorithms are able to compute better solutions than other recent heuristics for the problem scale well to large instances. 
The rest of this paper is organized as follows.  
In Section~\ref{sec:preliminaries}, we introduce basic concepts and describe relevant related work in more detail.
We present our main contributions in Section~\ref{s:main}. 
We implemented the techniques presented here in the graph partitioning framework KaHIP~\cite{kabapeE} (Karlsruhe High Quality Graph Partitioning). 
We present a summary of extensive experiments to evaluate algorithm performance in Section~\ref{sec:experiments}. The experiments indicate that our new integrated algorithm improves mapping quality over other state-of-the-art integrated and non-integrated mapping algorithms. 
For example, one configuration of our algorithm yields better solutions than the previous state-of-the-art in terms of mapping quality while being a factor 62 faster.
Compared to the currently fastest iterated multilevel mapping algorithm Scotch, we obtain 16\% better solutions while investing slightly more running time.
Most importantly, hierarchical multisection algorithms that take the system hierarchy into account for model creation improve the results of the overall process mapping significantly.

\section{Preliminaries}
\label{sec:preliminaries}

The communication requirements between the components of a set of processes in (some section of) an application can be represented by a weighted communication graph.
The underlying hardware topology can likewise be represented by a weighted graph, particularly a complete graph since any two physical processors can communicate with each other facilitated by the routing system.
This complete graph can be represented by a topology cost matrix reflecting the costs of routing along shortest or cheapest paths between physical processors.
Furthermore, it does not need to be explicitly expressed if the topology is organized as a regular hierarchy of components with fixed communication cost per message inside each level.
We tackle the problem of embedding a communication graph onto a topology graph under optimization criteria that we explain below. 
Unless otherwise mentioned, a processing element (PE) represents a core of a machine.

\subsection{Basic Concepts}
\label{subsec:basic_concepts}

Let $G=(V=\{0,\ldots, n-1\},E)$ be an \emph{undirected graph} with edge weights $\omega: E \to \MdR_{>0}$, vertex weights $c: V \to \MdR_{\geq 0}$, $n = |V|$, and $m = |E|$.
We generalize $c$ and $\omega$ functions to sets, such that $c(V') = \sum_{v\in V'}c(v)$ and $\omega(E') = \sum_{e\in E'}\omega(e)$.
Let $N(v) = \setGilt{u}{\set{v,u}\in E}$ denote the neighbors of a vertex $v$.
Let $I(v)$ denote the set of edges incident to $v$.
A graph $S=(V', E')$ is said to be a \emph{subgraph} of $G=(V, E)$ if $V' \subseteq V$ and $E' \subseteq E \cap (V' \times V')$. 
When $E' = E \cap (V' \times V')$, $S$ is an \emph{induced} subgraph.

The \emph{graph partitioning problem} (GPP) consists of assigning each node of $G$ to exactly one of $k$ distinct blocks respecting a balancing constraint in order to minimize the edge-cut.
More precisely, GPP partitions $V$ into $k$ blocks $V_1$,\ldots,$V_k$ (\ie $V_1\cup\cdots\cup V_k=V$ and $V_i\cap V_j=\emptyset$ for $i\neq j$), which is called a \emph{\mbox{$k$-partition}} of $G$.
The \emph{balancing constraint} demands that the sum of node weights in each block does not exceed a threshold associated with some allowed \emph{imbalance}~$\epsilon$. 
More specifically, $\forall i~\in~\{1,\ldots,k\} \gilt$ $c(V_i)\leq L_{\max}\Is \big\lceil(1+\epsilon) \frac{c(V)}{k} \big\rceil$.
Let a block $V_i$ be called \emph{$\lambda$-underloaded} if $|V_i| + \lambda \leq L_{\max}$ and \emph{overloaded} if $|V_i| > L_{\max}$.
The \emph{edge-cut} of a $k$-partition consists of the total weight of the edges crossing blocks, \ie $\sum_{i<j}\omega(E_{ij})$, where $E_{ij}\Is\setGilt{\set{u,v}\in E}{u\in V_i,v\in V_j}$. 
An abstract view of the partitioned graph is a \emph{quotient graph} $\mathcal{Q}$, in which nodes represent blocks and edges are induced by the connectivity between blocks. More precisely,  there is an edge in the quotient graph if there is an edge that runs between the blocks in the original, partitioned graph.
We call \emph{neighboring blocks} a pair of blocks connected to each other by an edge in the quotient graph.
If a node $v \in V_i$ has a neighbor $w \in V_j, i\neq j$, then it is called a \emph{boundary} node. 
Let $R(v)$ be the set of all blocks containing at least one element from $\{v\} \cup N(v)$.

Assume that we have $n$ processes and a topology containing $k$ PEs.
Let $\mathcal{C}\in \MdR^{n \times n}$ denote the communication matrix and let $\mathcal{D}\in \MdR^{k \times k}$ denote the (implicit) topology matrix or distance matrix.
In particular, $\mathcal{C}_{i,j}$ represents the required amount of communication between processes $i$ and $j$, while $\mathcal{D}_{x,y}$ represents the cost of each communication between PEs $x$ and $y$.
Hence, if processes $i$ and $j$ are respectively assigned to PEs $x$ and $y$, or vice-versa, the communication cost between $i$ and $j$ will be $\mathcal{C}_{i,j}\mathcal{D}_{x,y}$.
Throughout this work, we assume that $\mathcal{C}$ and $\mathcal{D}$ are symmetric -- otherwise one can create equivalent problems with symmetric inputs \cite{brandfass2013rank}.

In this work, we deal with topologies organized as homogeneous hierarchies, even though our algorithms could be extended to heterogeneous hierarchies in a
straightforward way. 
Let $\mathcal{S}=a_1: a_2: ...:a_\ell$ be a
sequence describing the hierarchy of a supercomputer. The sequence
should be interpreted as each processor having $a_1$ cores, each node
$a_2$ processors, each rack $a_3$ nodes, and so forth,
such that the total number of processors is $k=\Pi_{i=1}^{\ell}a_i$.
Let $D = d_1:d_2:\ldots:d_\ell$ be a sequence describing the communication cost inside each hierarchy level, meaning that two cores in the same processor communicate with cost $d_1$, two cores in the same node but in different processors communicate with cost $d_2$, two cores in the same rack but in different nodes communicate with cost $d_3$, and so forth.

Throughout the paper, we assume that the input communication matrix is already given as a graph $G_\mathcal{C}$, \ie no conversion of the matrix into a graph is necessary.
 More precisely, the graph representation is defined as $G_\mathcal{C}:=(\{1,\ldots, n\}, E[\mathcal{C}])$ where $E[\mathcal{C}] :=\{(u,v) \mid \mathcal{C}_{u,v} \not = 0\}$.
 In other words, $E[\mathcal{C}]$ is the edge set of the processes that need to communicate with each other. 
Note that the set contains forward and backward edges, and that the weight of each edge in the graph equals the corresponding entry in the communication matrix~$\mathcal{C}$.

Our \emph{main focus} in this work is the \emph{general process mapping problem}~(GPMP).
It consists of assigning each node of a given communication graph to a specific PE in a communication topology while respecting a balancing constraint (the same as in the graph partitioning problem above) in order to minimize the total communication costs.
Within the scope of this work, the number of nodes (processes)~$n$ in the communication graph is much larger than the number of PEs~$k$ in the topology graph which matches most real-world situations. 
Let the mapping function that maps a node onto its block be \mbox{$\Pi: \{1, \ldots, n\} \mapsto \{1, \ldots, k\}$}.
Hence, the  objective function of GPMP is to minimize $J(\mathcal{C},\mathcal{D}, \Pi) := \sum_{i,j} \mathcal{C}_{i, j}\mathcal{D}_{\Pi(i),\Pi(j)}$.
Many authors deal with the specific case in which $n=k$, resulting in the \emph{one-to-one process mapping problem} (OPMP), where each process~$i$ is assigned to a unique PE~$\Pi(i)$. 
Within the context of OPMP, searching for the inverse permutation instead, \ie assigning PE $x$ to node $\Pi^{-1}(x)$, results in the same problem since $\Pi$ is a bijection.

GPP and OPMP are both NP-hard problems \cite{Garey1974, SahniG76}. 
Since GPP and OPMP are special cases of GPMP, the latter is also \mbox{NP-hard}.
Hence, exact efficient algorithms to solve GPMP are very unlikely, which justifies the use of heuristics to obtain reasonably good solutions for real-world instances within a reasonable time.
Two of the most common methods to solve GPMP are the two-phase approach and the integrated approach.
In the \emph{two-phase} approach, GPMP is solved in two consecutive steps: 
(i) a heuristic for GPP is applied in the communication graph, obtaining a balanced $k$-partition; 
(ii) a heuristic for OPMP is used to map the blocks of the $k$-partition onto the topology of PEs.
On the other hand, the \emph{integrated} approach consists of tackling GPMP directly, \ie not decomposing the input problem into $k$ independent sub-problems first.

\subsection{Multilevel Approach}
\label{subsec:MultiLevel_Approach}

In this section, we characterize the multilevel approach within the scope of GPMP, although the same basic structure is extensible to many other problems, such as GPP.
Before describing the multilevel scheme, we need to define the terms contraction and uncontraction. 
\emph{Contracting} an edge $e=\set{u,v}$ consists of replacing the nodes $u$ and $v$ by a new node $x$ connected to the former neighbors of $u$ and $v$. 
We set $c(x)=c(u)+c(v)$ so that the weight of a node at each level is the sum of weights of the contracted nodes. 
If replacing edges of the form $\set{u,w}$, $\set{v,w}$ would generate two parallel edges $\set{x,w}$, a single edge with
$\omega(\set{x,w})=\omega(\set{u,w})+\omega(\set{v,w})$ is inserted.
The  \emph{uncontraction} of a node consists of undoing the contraction that gave rise to it.
In order to avoid tedious notation, $G$ will denote the current state of the graph, either before or after (un)contraction, unless we explicitly want to refer to 
different states of the graph.

A \emph{multilevel approach} to solve GPMP consists of three main phases.
In the \emph{contraction} (coarsening) phase, successive approximations of an original input graph are created. 
In particular, the first-level approximation is obtained directly through a contraction on the original graph, the second-level approximation is obtained through a contraction on the first-level approximation, and so forth.
Hence, each of these approximations conserves structural information about the input graph, but in different levels: from micro-structural (particular) information in the finest approximation to macro-structural (general) information in the coarsest approximation.
The contractions quickly reduce the size of the graph and stop as soon as it becomes sufficiently small to be partitioned and mapped by an expensive algorithm.
In the construction phase, an initial mapping is produced for the coarsest approximation of the input graph.
Due to the way we define contraction, every mapping of the coarsest level implies a corresponding mapping of the input graph with equal objective function and balance.
In the \emph{local improvement} (or uncoarsening) phase, we uncontract previously contracted nodes to go back through each level, from the coarsest approximation to the original graph. 
After each uncoarsening, local improvement algorithms move nodes between blocks in order to improve the objective function or balance.

\subsection{Related Work}
\label{subsec:related_work}

There has been an immense amount of research on GPP, and the reader is referred to~\cite{GPOverviewBook,SPPGPOverviewPaper,DBLP:reference/bdt/0003S19} for extensive material and references.
The most successful general-purpose methods to solve GPP for huge real-world graphs are based on the multilevel approach. 
The basic idea of this approach can be traced back to multigrid solvers for systems of linear equations~\cite{Sou35}, and its first application to GPP was by \citet{BarSim93}. 
The most commonly used formulation of the multilevel scheme was proposed by
\citeauthor{Hendrickson95}~\cite{Hendrickson95}.
Similar multilevel schemes are used for other graph partitioning problem formulations such as DAG partitioning\cite{DBLP:conf/gecco/MoreiraP018,DBLP:journals/siamsc/HerrmannOUKC19,DBLP:conf/wea/MoreiraPS17}, hypergraph partitioning~\cite{DBLP:conf/dac/KarypisK99,DBLP:conf/gecco/AndreS018}, graph clustering~\cite{DBLP:journals/corr/abs-0803-0476,DBLP:conf/aaim/DellingGSW09}, graph drawing~\cite{DBLP:journals/jgaa/Walshaw03,DBLP:journals/tvcg/MeyerhenkeN018} or the node separator problem~\cite{DBLP:conf/wea/Sanders016,DBLP:journals/coap/HagerHS18}.
Among the most successful multilevel software packages to solve GPP, we mention Jostle~\cite{walshaw2000mpm}, Metis~\cite{karypis1998fast}, Scotch~\cite{Scotch}, and KaHIP~\cite{kaHIPHomePage}.

Systems like KaHIP~\cite{kaHIPHomePage} and Metis~\cite{karypis1998fast} typically compute a $k$-partition on the coarsest level through a recursive bisection strategy or a direct $k$-way partitioning scheme. In recursive bisection, the graph is recursively divided into two blocks until the number of blocks is reached, \ie a bisection algorithm is used to split the graph into two blocks. 
More precisely, each bisection step itself uses a multilevel algorithm that stops as soon as the number of nodes is below a small threshold.
To obtain a bipartition in the coarsest level, KaHIP uses the \emph{greedy graph growing} algorithm.
In KaHIP, if $k$ is not even, the graph gets split into two blocks, $V_1$ and $V_2$, such that $c(V_1) \leq \big\lfloor \frac{k}{2}\big\rfloor L_{\max}$, $c(V_2) \leq \big\lceil \frac{k}{2}\big\rceil L_{\max}$.
Block $V_1$ will be recursively  partitioned in $\lfloor \frac{p}{2} \rfloor$ blocks and block $V_2$ will be recursively partitioned in $\lceil \frac{p}{2} \rceil$ blocks.

In addition to GPP, Jostle and Scotch can also solve GPMP.
Jostle integrates local search into a multilevel scheme to partition the model of
computation and communication. 
In this scheme, it solves the problem on the coarsest level and afterwards performs refinements based on the user-supplied network communication model. 
On the other hand, Scotch performs dual recursive bipartitioning to compute a mapping.
More precisely, it starts the recursion considering all given processes and PEs. 
At each recursion level, it bipartitions the communication graph and also the distance graph with a graph bipartitioning algorithm. 
The first (resp., second) block of the communication graph is then assigned to the first (resp., second) block of the distance graph. 
The recursion proceeds until the distance graph only contains one vertex.

There is likewise a large literature on OPMP, often in the context of scientific applications using the \emph{Message Passing Interface} (MPI).
\citet{Hatazaki98} was among the first authors to propose graph partitioning to solve the MPI process mapping of a virtual unweighted topology onto a hardware topology organized in modules and sub-modules.
\citeauthor{Traff02:topology}~\cite{Traff02:topology} used a similar approach to implement one of the first non-trivial mappings designed for the NEC SX-series of parallel vector computers.  
\citet{mercier2009towards} and later \citet{MercierJeannot11} simplified the mapping problem to ignore the whole network topology except that inside each node.
They also investigated multiple placement policies to enhance overall system performance.  
\citet{YuChungMoreira06} discussed and implemented graph embedding heuristics for the BlueGene $3d$ torus systems.  
\citet{HoeflerSnir11} optimized instead the congestion of the mapping, additionally providing a proof that this problem is NP-complete, a heuristic to solve it, and an experimental evaluation based on application data from the Florida Sparse Matrix Collection.  
Routing-aware mapping heuristics taking the hierarchy of specific hardware topologies into account were discussed in~\cite{Abdel-GawadThottethodiBhatele14}.
\citet{Vogelstein15} concentrated on solving OPMP. 
They proposed a gradient--based heuristic that involves solving assignment problems and gave experimental evidence for better solution quality and speed compared to other heuristics.

\citet{muller2013optimale} proposed a greedy construction method to obtain an initial permutation for OPMP.
The method roughly works as follows:
Initially compute the total communication volume for each process and also the sum of distances from each core to all the others.
Note that this corresponds to the weighted degrees of the nodes in the communication and distance models, respectively. 
Afterwards, the process with the largest communication volume is assigned to the core with the smallest total distance. 
To build a complete assignment, the algorithm proceeds by looking at unassigned processes and cores. 
For each of the unassigned processes, the communication load to already assigned vertices is computed. 
For each core, the total distance to already assigned cores is computed. 
The process with the largest communication sum is assigned to the core with the smallest distance sum. \citet{glantzMapping2015} noted that the algorithm does not link the choices for the vertices and cores and hence propose a modification of this algorithm called \emph{GreedyAllC} (the best algorithm in \cite{glantzMapping2015}). 
GreedyAllC links the mapping choices by scaling the distance with the amount of communication to be done. 
The algorithm has the same asymptotic complexity and memory requirements as the algorithm by \citeauthor{muller2013optimale}. 

\citet{heider1972computationally} proposed a method to improve an already given solution for OPMP. 
The method repeatedly tries to perform swaps in the assignment. 
To do so, the author defines a pair-exchange neighborhood $N(\Pi)$ that contains all permutations that can be reached by swapping two elements in $\Pi$. 
Here, swapping two elements means that $\Pi^{-1}(i)$ will be assigned to processor $j$ and $\Pi^{-1}(j)$ will be assigned to processor $i$ after the swap is done. 
The algorithm then looks at the neighborhood in a cyclic manner. 
More precisely, in each step the current pair $(i,j)$ is updated to $(i,j+1)$ if $j<n$, to $(i+1,i+2)$ if $j=n$ and $i < n-1$, and lastly to $(1,2)$ if $j=n$ and $i = n-1$. 
A swap is performed if it yields a positive gain, \ie the swap reduces the objective. 
The overall runtime of the algorithm is $O(n^3)$. 
We denote the~search~space~with~$N^2$.
To reduce the runtime, \citet{brandfass2013rank} introduced a couple of modifications. 
First, only symmetric inputs are considered. 
If the input is not symmetric, it is substituted by a symmetric one such that the output of the algorithm remains the same. 
Second, pairs $(i,j)$ for which the objective cannot change are not considered.
For example, if two processes reside on the same compute node, swapping them will not change the objective. 
Third, the authors partition the neighborhood search space into $s$ consecutive index blocks and only perform swaps inside those blocks. 
This reduces the number of possible pairs from $O(n^2)$ to $O(ns)$ overall pairs. 
We denote the search space with $\mathcal{N}_p$ (\emph{pruned neighborhood}).
In addition, the authors use the method of \citet{muller2013optimale} to compute an initial solution.

\citet{schulz2017better} tackled the GPMP using a two-phase approach.
First, the graph is partitioned using KaHIP (which uses recursive bisection). The quotient graph (communication graph) is then the input to OPMP.
This is solved using a construction algorithm called hierarchy top down and a variation of the refinement method proposed by \citet{brandfass2013rank}.
\emph{Hierarchy top down} consists of a perfectly balanced multisection partitioning algorithm that partitions the communication graph recursively into blocks specified by the given hierarchy.
The applied refinement method is based on the local search proposed by Brandfass \etal \cite{brandfass2013rank}, but with alternative schemes of swapping neighborhoods and better data structures to improve performance.
Each of these schemes is represented by $N_{\mathcal{C}}^{d}$, in which a swap of two blocks is only allowed if their theoretical distance in the communication graph $G_\mathcal{C}$ does not exceed $d$. 
For instance, in the simplest swapping neighborhood $N_{\mathcal{C}}^{1}$, assignments can only be switched if the two blocks are connected by an edge in $G_\mathcal{C}$. 
This definition implies a sequence of neighborhoods increasing in size $N^1_\mathcal{C} \subseteq N^2_\mathcal{C} \subseteq \ldots \subseteq N^n_\mathcal{C} =  N^2$ where $N^2$ is the largest neighborhood used by \citet{brandfass2013rank}.
\citet{schulz2017better} experimentally showed that $N^{10}_\mathcal{C}$ is an adequate choice to obtain good solutions with a moderate running time.

The OPMP algorithm proposed by \citet{DBLP:conf/icpp/GlantzPM18} requires that the
hardware topology is a partial cube, i. e. an isometric subgraph of a
hypercube. This requirement allows to label (i) the PEs as well as (ii)
the nodes of the application graph $G$ with meaningful bit-strings along
convex cuts. These bit-strings facilitate (i) the fast computation of
distances between PEs and (ii) an effective hierarchical local search
method to improve the mapping induced by the labels.

Subsequently, \citet{GlobalMultisection} modified the graph partitioning step such that is already using  \emph{hierarchical multisection} itself.
This yields better communication graphs for the second OPMP mapping step.
In particular, all the processes are partitioned among all the available data centers, then the processes of each data center are partitioned among the servers contained in it, and so forth. 
However, this approach restricts the movement of nodes to the module in which the local searches of KaHIP are operating at each moment.
In~our integrated approach, instead, we adapt the local searches of KaHIP for the objective function of GPMP, which allows them to freely move nodes between any two blocks at any time.
After the partitioning step, \citet{GlobalMultisection} test different OPMP algorithms and demonstrate that an \emph{identity} mapping followed by an $ N^{10}_\mathcal{C}$ local search provides a good balance between solution quality and runtime performance.
Here we use a \emph{multisection} setup followed by an \emph{identity} mapping to compute initial solutions in the coarsest level of our multilevel approach, and give more details in Section \ref{subsec:Initial Solution}.

\section{High-Quality Multilevel General Process Mapping}
\label{s:main}
\label{sec:integrated_mapping_approach}

\begin{figure}[b]
\centering
\includegraphics[width=0.65\columnwidth]{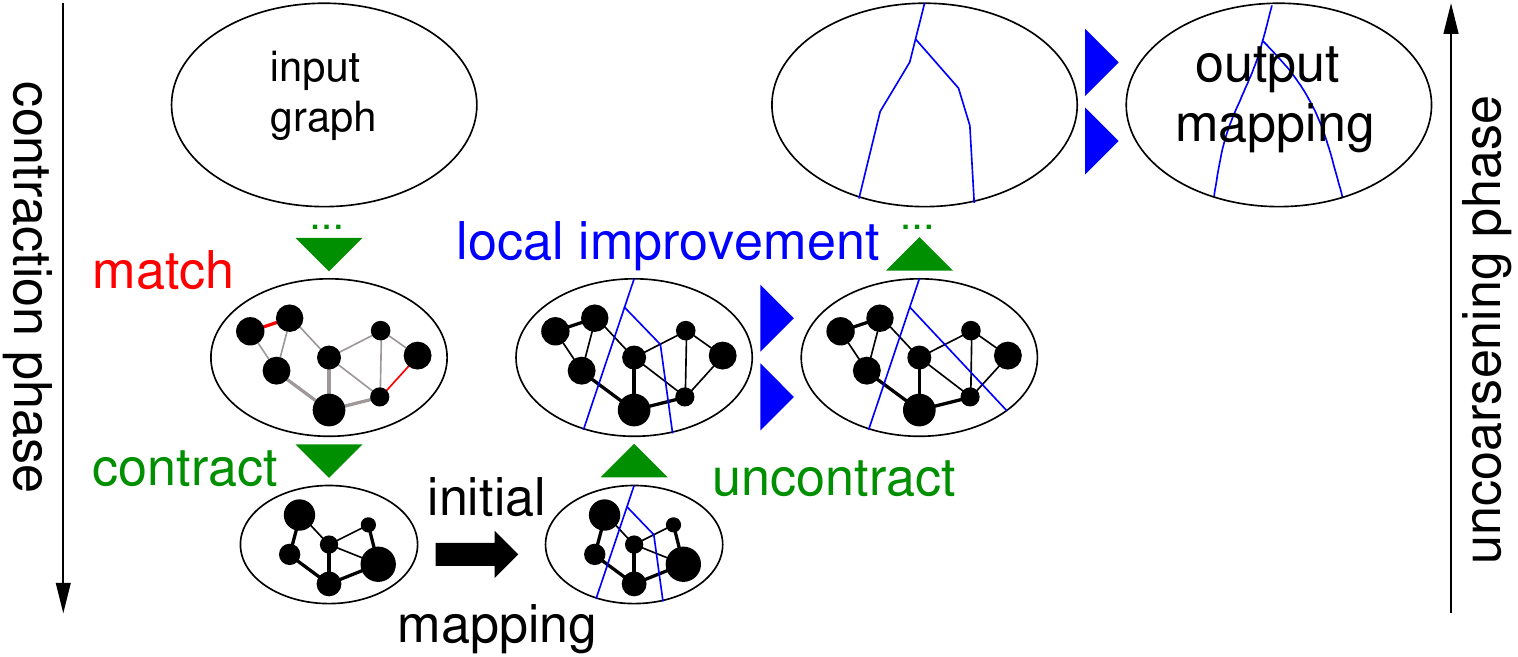}
\caption{Multilevel scheme used to solve GPMP (Figure from \cite{kaffpa})}\label{fig:multilevel}
\end{figure}

We engineered all the components of a multilevel algorithm to solve GPMP in an \emph{integrated} way, as illustrated in Figure \ref{fig:multilevel}.
In this section, we present our algorithmic contributions and discuss each of their components. 
This includes coarsening-uncoarsening schemes, methods to obtain initial solutions, local refinement methods, and additional tools to explore trade-offs in memory usage and performance.

\subsection{Coarsening}
\label{subsec:Contraction}
\ifFull
\else
We use a matching-based coarsening scheme.
The \emph{matching--based} coarsening is the most common choice in multilevel partitioning algorithms due to its simplicity, speed, and generality. 
It has two consecutive steps: An edge rating function and a matching algorithm. 
Based on local information, the \emph{edge rating function} scores each edge to estimate the benefit of contracting it.
We employ the same edge rating function $\text{exp*}(e)=  \omega(e)/(d(u)d(v))$ as used in \citet{kaffpa}.
Then, the \emph{matching algorithm} obtains a maximal match to maximize the sum of the ratings of the contracted edges. 
As in \cite{kaffpa}, we computed matchings with the \emph{Global Paths Algorithm} \cite{kaffpa}, which is a $\frac{1}{2}$-approximate algorithm.

\fi{}

\subsection{Initial Solution Algorithms}
\label{subsec:Initial Solution}

We compute the initial mapping using a two-phase approach.
To solve GPP, we compare two multilevel recursive bisection algorithms:
(i)~\emph{standard bisection} setup, in which we perform a recursive bisection to obtain $k$ blocks;
(ii)~\emph{multisection} setup, in which we perform recursive bisections throughout the hierarchical structure of PEs.
To construct a solution for OPMP, we apply two different construction methods: 
(i)~\emph{identity}, which automatically assigns each block to the PE with the same ID; 
(ii)~\emph{hierarchy top down}, which partitions the set of blocks throughout the hierarchical structure of PEs.
To refine the OPMP solution, we perform an $N^{10}_\mathcal{C}$ swap neighborhood local search.
Hence, the resulting map~$\Pi$ of nodes to PEs becomes our initial GPMP solution.

Our \emph{standard bisection} setup for initial partition  corresponds to the initial partition step in KaHIP.
Moreover, it is a canonical choice to produce initial solutions in multilevel schemes tackling GPP.
On the other hand, the \emph{multisection} setup draws inspiration from the scheme used in~\cite{GlobalMultisection}.
It is an attempt to specialize the initial partition for the particular case tackled in this paper: a regularly hierarchical distribution of PEs in which the communication cost between two processes (nodes) highly depends on the hierarchy level shared by their corresponding PEs (blocks).
Particularly, we apply a recursive partitioning scheme that splits all the nodes in $a_\ell$ blocks, then splits the nodes in each block in $a_{\ell-1}$ sub-blocks, then splits the nodes in each sub-blocks in $a_{\ell-2}$ sub-sub-blocks, and so forth.
Observing that the communication costs decrease as the communicating processes share lower hierarchy levels, the multisection approach implies a hierarchy of sub-problems that directly reflects the problem cost hierarchy.

In both setups of the partitioning step, we recursively assign consecutive IDs to blocks throughout the process in order to maintain locality.
Moreover, the PEs belonging to each hierarchy module are labeled with consecutive IDs, which also promotes locality. 
Then, the \emph{identity} method is a fast way to construct a solution for OPMP taking advantage of this locality: it assigns each block $V_i$ to the PE with the ID $i$.
Note, the \emph{standard bisection} setup conveniently combines with the identity mapping approach when $k$ is a power of $2$ since the recursive bisections will be automatically performed throughout the hierarchical topology.
For an analogous reason, the \emph{multisection} setup is a good algorithm to create a coarse model to be mapped by the \emph{identity} mapping approach independently of~$k$.
The \emph{hierarchy top down}~\cite{schulz2017better} is a more general approach to construct solutions for OPMP when the PEs are hierarchically organized.
Its mechanism is similar to the idea of multisection throughout the hierarchy.

\subsection{Uncoarsening}
\label{subsec:uncoarsening}

After obtaining an initial solution for GPMP at the coarsest level, we apply a sequence of four local refinement methods to move nodes between blocks (which are already associated to unique PEs).
Then, we undo each of the contractions performed previously, from the coarsest graph until the original input graph.
After each uncoarsening step, we repeat our four local refinement methods.
The refinements run in a specific order based on their characteristics.
First, a \emph{quotient graph refinement} exhaustively tries to improve solution quality and eliminate imbalance by moving nodes between each pair of blocks connected by an edge in the quotient graph.
Second, a \emph{$k$-way Fiduccia-Mattheyses (FM) algorithm~\cite{fiduccia1982lth} refinement} greedily goes through the boundary nodes trying to relocate them with a more global perspective in order to improve the mapping.
Third, a \emph{label propagation refinement} randomly visits all nodes and moves each one to the most appropriate block while not decreasing the objective.
Finally, a \emph{multi-try FM refinement} is exhaustively applied in rounds with random starting points throughout the graph in order to escape local optima as many times as possible.
Before explaining the local search algorithms, we introduce the notion of \emph{gain} for GPMP.

\subparagraph*{Gain.}

All our refinement methods are based on the concept of \emph{gain}.
Equation~(\ref{eq:local_obj_func}) defines $\Psi_b(v)$ as the \emph{partial} contribution of a node $v$ to the objective function $J(\mathcal{C},\mathcal{D}, \Pi)$ in case $v$ is assigned to the PE $b$.
More precisely, $\Psi_b(v)$ represents the total cost of the communications involving~$v$ if $\Pi(v)=b$ and the neighbors of $v$ remain assigned to their current PEs.
Based on this definition, Equation (\ref{eq:gain}) defines the \emph{gain}~$g_b(v)$, which represents the value that will be subtracted from $J(\mathcal{C},\mathcal{D}, \Pi)$ if a node $v$ is moved from its current PE $\Pi(v)$ to PE $b$.
\begin{align}
    \Psi_b(v) & \Is \sum_{\set{v,u} \in I(v)} {\mathcal{C}_{v,u} \mathcal{D}_{b,\Pi(u)}}
    \label{eq:local_obj_func} \\
    g_b(v) & \Is \Psi_{\Pi(v)}(v) - \Psi_b(v)
    \label{eq:gain} 
\end{align}
Definition (\ref{eq:gain}) implies $g_{\Pi(v)}(v) \equiv 0$. Observe that a positive (resp., negative) gain indicates improvement (resp., worsening) of the solution.
Computing the gains of~$v$ to all blocks in $R(v)$ costs $O\big(|R(v)||I(v)|\big) = O\big(|I(v)|^2\big)$.
For comparison purposes, the computation of the same corresponding gains in the context of GPP and edge-cut objective function costs $O\big(|I(v)|\big)$.

\subparagraph*{Quotient Graph Refinement.}

We implemented an adapted version of the \emph{quotient graph refinement} \cite{kaffpa} to incorporate our definition of gains.
Within this refinement, we visit each pair of neighboring blocks in the quotient graph $\mathcal{Q}$ underlying the current $k$-partition. 
Then we apply an FM algorithm~\cite{fiduccia1982lth} to move nodes between the two currently visited blocks, keeping two respective gain--based priority queues of eligible nodes.
Each queue is randomly initialized with the boundary in its corresponding block.
After a node is moved (which can only happen once during an execution of the local search), its unmoved neighbors become eligible.
We employ the \emph{TopGain} scheme to select the block from which the next node will be moved and the \emph{active block scheduling}, both proposed by \citet{kaffpa}.
This refinement method includes strategies to favor the removal of nodes from overloaded blocks and to escape from local optima.

\subparagraph*{K-Way FM Refinement.}
\label{subsec:K-Way_FM_Refinement}

Our $k$-way FM refinement was adapted from the implementation in~\cite{kaffpa}.
Unlike the quotient graph refinement, the $k$-way FM does not restrict the movement of a node to a certain pair of blocks, but performs global-aware movement choices.
Our implementation of $k$-way FM uses only one gain--based priority queue $P$, which is initialized with the \emph{complete} partition boundary in a random order.   
Then, the local search repeatedly looks for the highest-gain node $v$ and moves it to the best $c(v)$-underloaded neighboring block.
When a node is moved, we insert in $P$ all its neighbors that were not in $P$ and have not been moved yet.
The $k$-way local search stops if $P$ is empty (\ie each node was moved once) or when a stopping criterion based on a random-walk model described in~\cite{kaffpa} applies.
To escape from local optima, this refinement allows some movements with negative gain or to blocks that are not  $c(v)$-underloaded.
Afterwards local search is rolled back to the lowest cut fulfilling the balance criterion that occurred.

\subparagraph*{Label Propagation Refinement.}

We propose a local search inspired by \emph{label propagation} \cite{labelpropagationclustering}. 
The algorithm works in rounds.
In each round, the algorithm visits all nodes in a random order, starting with the labels being the current assignment of nodes to blocks.
When a node $v$ is visited, it is moved to the $c(v)$-underloaded neighboring block  with highest positive gain.
We consider only $c(v)$-underloaded blocks since this ensures that the target block is not overloaded when the node is moved there.
Ties are broken randomly and a 0-gain neighboring block can be occasionally chosen with $50\%$ probability if there is no neighboring $c(v)$-underloaded block with positive gain.
We perform at most~$\ell$ rounds of the algorithm, where $\ell$ is a tuning parameter.

\subparagraph*{Multi-Try FM Refinement.}
\label{subsec:Multi-Try_FM_Refinement}

We also adapted our gain concept to a localized variant of the $k$-way local search algorithm similar to that proposed in \cite{kaffpa} under the name of \emph{multi-try} FM.
Instead of being initialized with all boundary nodes, as in $k$-way FM, multi-try FM is repeatedly initialized with a single boundary node. 
This introduces a higher diversification to the search since it is not restricted to movements in boundary nodes with global largest gain.
As a result, this local search can escape local optima more easily than \emph{$k$-way} FM.

{}

\subsection{Additional Techniques}
\label{subsec:additional_techniques}

We implemented some techniques that yield a memory vs running time trade-off.
In this section, we explain our approaches to deal with the topology matrix and the recomputation of gains.

\subparagraph*{Implicit Distance Matrix.}

When the topology matrix $\mathcal{D}$ is stored in memory, the time complexity to obtain the distance between a pair of PEs is $O(1)$, but this requires $O(k^2)$ space. 
From now on, we refer to the algorithm explicitly keeping $\mathcal{D}$ in memory as \emph{matrix--based} approach.
We implement three alternative approaches to save memory by exploiting the fact that our topology matrix is a hierarchy and the IDs of PEs in each of the hierarchy modules are sequential.
For simplification reasons, we call these approaches: (i) \emph{division--based}; (ii) \emph{stored division--based}; and (iii) \emph{binary notation--based}.

\begin{figure}[b]
\centering

\includegraphics[width=0.7\textwidth]{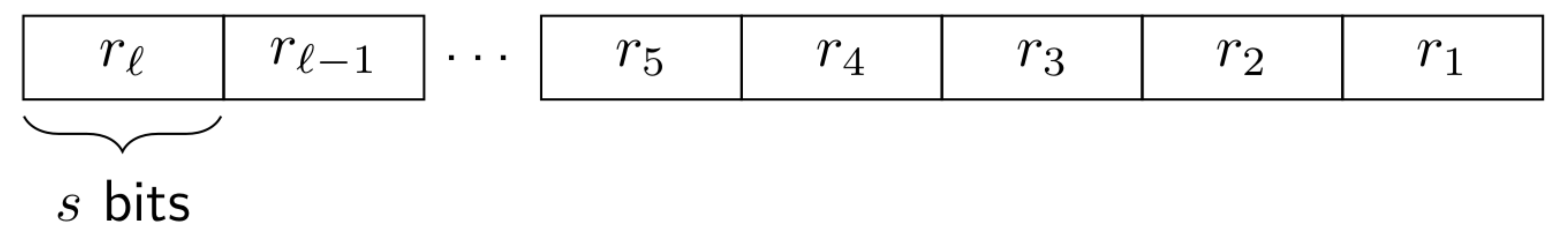}

\caption{Section structure of the binary number used to represent PE~$b$.} \label{fig:bina_repr}
\end{figure}

In the \emph{division--based} approach, we perform $O(\ell)$ successive integer divisions and comparisons in the ID of two PEs whenever we need to find out their distance. Here, $\ell$ is the number of levels in the system hierarchy. 
As a preprocessing step executed only once, we create a vector $h=\Big\{k\Big/{\prod_{t=1}^{\ell} a_{t}}, \; k\Big/{\prod_{t=2}^{\ell} a_{t}},\; \ldots,\; k\Big/{a_{\ell}}\Big\}$.
To find the distance between PEs $b$ and $b^\prime$ with $b \neq b^\prime$, we loop through the hierarchy layers from $i=\ell$ to $i=1$.
In each iteration, we perform the integer division of $b$ and $b^\prime$ by $h_{i}$.
Whenever the division results differ, then we break the loop and return $\mathcal{D}_{b,b^\prime}=d_{i}$.
Summarizing, this approach does not require any additional memory other than a vector with $O(\ell)$ integers and has complexity $O(\ell)$.

The \emph{stored division--based} approach works in a similar way as the \emph{division--based} one.
The only difference is that we avoid repetitive integer divisions of IDs by elements of $h$ by storing the results of all possible divisions in a preprocessing step executed only once.
Although we still need $O(\ell)$ running time to perform comparisons in order to obtain the distance between a pair of PEs, the constant factors involved are much lower.
This improvement in running time comes in the cost of additional $O(k\ell)$ memory space.

The \emph{binary notation--based} approach is a more compact way of decomposing the IDs of PEs.
Instead of storing $\ell$ numbers for each PE, we keep in memory a single binary number per PE.
This binary number $r$ consists of $\ell$ sections $r_i$, each containing $s=\big\lceil \log_{2}\big(\max_{1\leq t \leq \ell}(a_t) \big) \big\rceil$ bits (see Figure \ref{fig:bina_repr}).
To describe the construction of $r$ for a PE~$b$, let a variable~$t$ be initialized as $t=b$.
Then, we loop through the hierarchy layers, from $i=1$ to $i=\ell$.
In each iteration~$i$, $r_{i}$ receives the remainder of the division of $t$ by $a_{i}$ and, then, $t$ is updated to store the integer quotient of $t$ by $a_{i}$.
Afterwards, it is possible to precisely locate $b$ at the hierarchy by sweeping the sections of $r$ from $r_{\ell}$ to $r_1$.
In particular, $r_{\ell}$ specifies its data center, $r_{\ell-1}$ specifies its server among those belonging to its data center, and so forth.
Obtaining the distance between distinct PEs $b$ and $b^\prime$ is equivalent to finding which section $r_i$ contains the leftmost nonzero bit in the result of the bit-wise operation XOR($b$,$b^\prime$).
The running-time complexity of finding the section of the leftmost nonzero bit is $O(\log(\ell))$.
Furthermore, current processors often implement a \emph{count leading zeros} (CLZ) operation in hardware which allows the identification of the leftmost nonzero bit in $O(1)$ time, under the assumption that the size $\log r = O(\log k)$ of the
   binary numbers is smaller than the size of a machine word.

\subparagraph*{Delta-Gain Updates.}

Our local searches frequently need to compute \emph{gains} involved in the movement of nodes.
A \emph{base approach} to check these gains consists of computing them from scratch whenever they are needed, which can yield many gain recomputations.
For this reason, we implement a technique to save running time  called \emph{delta-gain updates} \cite{schlag2016k}.

\tikzset{Nodo/.style={circle,fill=gray!25, minimum size=7mm}}

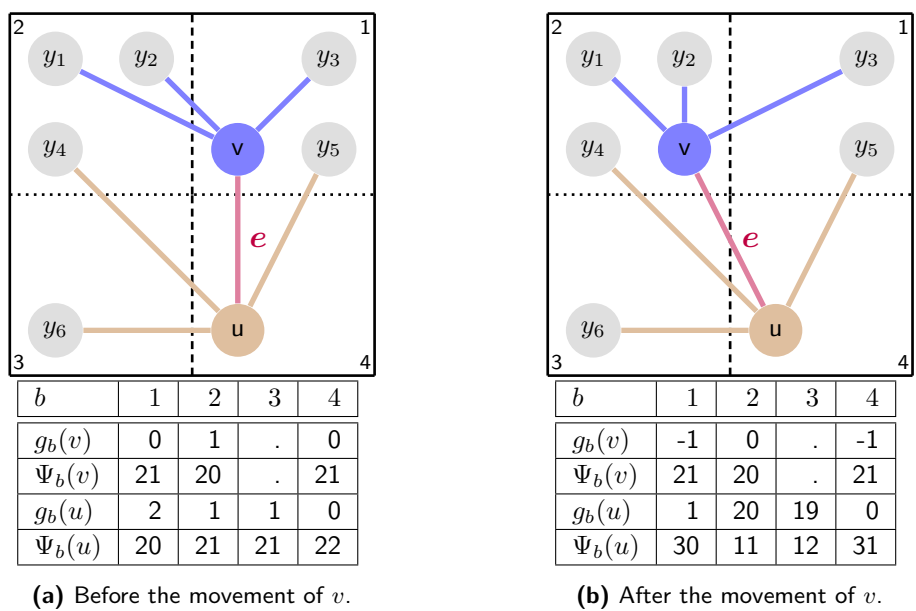
\begin{figure}[!b]
\captionsetup[subfigure]{justification=centering}
	\begin{subfigure}[b]{0.5\textwidth}
		\centering
\begin{tikzpicture}
		[xscale=0.06, yscale=0.06,auto=center]

	\node[color=black] at (38,37) {\fontsize{8}{8}\selectfont 1};
	\node[color=black] at (-38,37) {\fontsize{8}{8}\selectfont 2};
	\node[color=black] at (-38,-37) {\fontsize{8}{8}\selectfont 3};
	\node[color=black] at (38,-37) {\fontsize{8}{8}\selectfont 4};

	\draw[color=black,line width=1,   dotted] (-40,0) -- (40,0) ;

	\draw[color=black,line width=1, densely dashed] (0,-40) -- (0,0) ; 
	\draw[color=black,line width=1, densely dashed] (0,0) -- (0,40) ;

	\draw[color=black,line width=1, solid] (-40,-40) -- (-40,40) ;
		
	\draw[color=black,line width=1, solid] (40,-40) -- (40,40) ;
		
	\draw[color=black,line width=1, solid] (-40,-40) -- (40,-40) ;
		
	\draw[color=black,line width=1, solid] (-40,40) -- (40,40) ;

	\node[Nodo] (y1) at (-30, 30) {$y_1$};
	\node[Nodo] (y2) at (-10, 30) {$y_2$};
	\node[Nodo] (y4) at (-30, 10) {$y_4$};
		
	\node[Nodo] (y3) at (30, 30) {$y_3$};
	\node[Nodo] (y5) at (30, 10) {$y_5$};
		
	\node[Nodo] (y6) at (-30, -30) {$y_6$};

    \node[circle,fill=blue!50, minimum size=7mm] (my_v) at ( 10, 10) {v};

    \node[circle,fill=brown!50, minimum size=7mm] (my_u) at ( 10, -30) {u};

	\draw[line width=2, color=purple!50] (my_v) -- node[text=purple,right] {\fontsize{12}{12}\selectfont $\boldsymbol{e}$} (my_u);
	\draw[line width=2, color=blue!50] (my_v) -- (y1);
	\draw[line width=2, color=blue!50] (my_v) -- (y2);
	\draw[line width=2, color=blue!50] (my_v) -- (y3);
	\draw[line width=2, color=brown!50] (my_u) -- (y4);
	\draw[line width=2, color=brown!50] (my_u) -- (y5);
	\draw[line width=2, color=brown!50] (my_u) -- (y6);

		\end{tikzpicture}
		
\begin{tabular}{|l|r|r|r|r|}
\hline
 ${b}$ &  ${1}$ &  ${2}$ &  ${3}$ &  ${4}$ \\ \hline \hline
 ${g_b(v)}$ & 0 & {1} & . & 0  \\ \hline
 ${\Psi_b(v)}$ & 21 & 20 & . & 21  \\ \hline
 ${g_b(u)}$  & {2}  & {1}  & {1}  & 0 \\ \hline
 ${\Psi_b(u)}$  &  20 & 21  & 21  & 22 \\ \hline
\end{tabular}
		\caption{Before the movement of $v$.}
		\label{fig:discrA}
	\end{subfigure}
	\begin{subfigure}[b]{0.5\textwidth}
		\centering
\begin{tikzpicture}
		[xscale=0.06, yscale=0.06,auto=center]

	\node[color=black] at (38,37) {\fontsize{8}{8}\selectfont 1};
	\node[color=black] at (-38,37) {\fontsize{8}{8}\selectfont 2};
	\node[color=black] at (-38,-37) {\fontsize{8}{8}\selectfont 3};
	\node[color=black] at (38,-37) {\fontsize{8}{8}\selectfont 4};

	\draw[color=black,line width=1,   dotted] (-40,0) -- (40,0) ;

	\draw[color=black,line width=1, densely dashed] (0,-40) -- (0,0) ; 
	\draw[color=black,line width=1, densely dashed] (0,0) -- (0,40) ;

	\draw[color=black,line width=1, solid] (-40,-40) -- (-40,40) ;
		
	\draw[color=black,line width=1, solid] (40,-40) -- (40,40) ;
		
	\draw[color=black,line width=1, solid] (-40,-40) -- (40,-40) ;
		
	\draw[color=black,line width=1, solid] (-40,40) -- (40,40) ;

	\node[Nodo] (y1) at (-30, 30) {$y_1$};
	\node[Nodo] (y2) at (-10, 30) {$y_2$};
	\node[Nodo] (y4) at (-30, 10) {$y_4$};
		
	\node[Nodo] (y3) at (30, 30) {$y_3$};
	\node[Nodo] (y5) at (30, 10) {$y_5$};
		
	\node[Nodo] (y6) at (-30, -30) {$y_6$};

    \node[circle,fill=blue!50, minimum size=7mm] (my_v) at ( -10, 10) {v};

    \node[circle,fill=brown!50, minimum size=7mm] (my_u) at ( 10, -30) {u};

	\draw[line width=2, color=purple!50] (my_v) -- node[text=purple,right] {\fontsize{12}{12}\selectfont $\boldsymbol{e}$} (my_u);
	\draw[line width=2, color=blue!50] (my_v) -- (y1);
	\draw[line width=2, color=blue!50] (my_v) -- (y2);
	\draw[line width=2, color=blue!50] (my_v) -- (y3);
	\draw[line width=2, color=brown!50] (my_u) -- (y4);
	\draw[line width=2, color=brown!50] (my_u) -- (y5);
	\draw[line width=2, color=brown!50] (my_u) -- (y6);

		\end{tikzpicture}
		
\begin{tabular}{|l|r|r|r|r|}
\hline
 ${b}$ &  ${1}$ &  ${2}$ &  ${3}$ &  ${4}$  \\ \hline \hline
 ${g_b(v)}$ & {-1} & 0 & . & {-1}  \\ \hline
 ${\Psi_b(v)}$ & 21 & 20 & . & 21  \\ \hline
 ${g_b(u)}$  & {1}  & {20}  & {19}  & 0 \\ \hline
 ${\Psi_b(u)}$  &  30 & 11  & 12  & 31 \\ \hline
\end{tabular}
		\caption{After the movement of $v$.}
		\label{fig:discrB}
	\end{subfigure}
	\caption{The diagrams in (a) and (b) represent eight nodes embedded in a hierarchy described by $\mathcal{S}=2:2$ and $D=1:10$ before and after the movement of node $v$ from PE 1 to PE 2.
	The dashed line represents the communication channel of cost 10, the dotted lines represent the communication channels of cost 1, and the solid lines between nodes represent edges with weight 1.
	The table below each diagram shows the gains and partial objective functions of $v$ and $u$ for each respective configuration.}
	\label{fig:delta_update}
\end{figure}

In \emph{delta-gain updates}, we store a vector of length $O(|R(v)|)=O(|I(v)|)$ for each node~$v$.
In this vector, we keep the gains $g_{b}(v)$ for all PEs $b$ containing neighbors of $v$.
Additionally, we store an $n$-sized vector $h$ to keep flags that indicate whether a node has up-to-date gains in memory.
Asymptotically speaking, these vectors represent $O(n+m)$ extra memory.
Each flag is initialized with an inactive seed and is considered active if its value equals the number of uncoarsening steps performed so far.
When we need to check a gain of some node $v$, we look at $h_v$ to verify if the gains of $v$ are up-to-date.
If they are not, we compute all gains $g_{b}(v)$ from scratch, which costs $O\big(|I(v)|^2\big)$, and activate $h_v$.
Otherwise, we just access the required gain from memory in $O(1)$ time.

If a node $v$ moves from its current PE to another one, we have to update all delta gains of $v$ and $u \in N(v)$ with $h_u$ being active.
We use Figure \ref{fig:delta_update} as an illustrative example to explain how to update these delta gains.
Assume that $h_v$ and $h_u$ are active and $v$ moves from PE 1 to PE 2 during some local refinement.
After this movement, we should change the delta gains of $u$ and $v$ in memory.
For $v$, it suffices to subtract $g_2(v)$ from all other gains of $v$ and then set $g_2(v)$ to 0.
\begin{wraptable}{r}{0.45\textwidth}
	\centering
        \small
 	\caption{Benchmark instance properties.}
        \vspace*{-.25cm}
 	\label{tab:test_instances_walshaw}
	\begin{tabular}{| l | r | r | }
			\hline
		 	Graph & $n$& $m$\\
		 	\hline \hline
		 	 \multicolumn{3}{|c|}{Tuning Graphs}\\
			\hline

		 	  ecology2                         & $\approx$1.0M  & \numprint{1997996}\\
		 	  G3\_circuit                                & $\approx$1.6M & \numprint{3037674}\\
		 	  fe\_rotor                                 & \numprint{99617} & \numprint{662431}\\
		 	  598a                                           & \numprint{110971} & \numprint{741934}\\
		 	  del22                                        & $\approx$4.2M & $\approx$12.6M\\
		 	  rgg22                                       & $\approx$4.2M & $\approx$30.4M\\

                          \hline
		 	 \multicolumn{3}{|c|}{UF Graphs}\\
			\hline

		 	  cop20k\_A                                     & \numprint{99843}  & \numprint{1262244}\\
		 	  2cubes\_sphere                                & \numprint{101492} & \numprint{772886}\\
		 	  thermomech\_TC                                 & \numprint{102158} & \numprint{304700}\\
		 	  cfd2                                           & \numprint{123440} & \numprint{1482229}\\
		 	  boneS01                                        & \numprint{127224} & \numprint{3293964}\\
		 	  Dubcova3                                       & \numprint{146689} & \numprint{1744980}\\
		 	  bmwcra\_1                                      & \numprint{148770} & \numprint{5247616}\\
		 	  G2\_circuit                                    & \numprint{150102} & \numprint{288286} \\
		 	  shipsec5                                       & \numprint{179860} & \numprint{4966618}\\
		 	  cont-300                                       & \numprint{180895} & \numprint{448799}  \\
		 	
                          \hline
		 	  \multicolumn{3}{|c|}{ Large Walshaw Graphs}  \\
		 	
                          \hline
		 	  598a                                           & \numprint{110971} & \numprint{741934}   \\
		 	  fe\_ocean                                      & \numprint{143437} & \numprint{409593}   \\
		 	  144                                            & \numprint{144649} & \numprint{1074393}  \\
		 	  wave                                           & \numprint{156317} & \numprint{1059331} \\
		 	  m14b                                           & \numprint{214765} & \numprint{1679018}  \\
		 	  auto                                           & \numprint{448695} & \numprint{3314611}  \\
		 	
                          \hline
		 	   \multicolumn{3}{|c|}{ Large Other Graphs}\\
		 	
                          \hline
		 	  del23                                          & $\approx$8.4M     & $\approx$25.2M \\
		 	  del24                                          & $\approx$16.7M    & $\approx$50.3M \\
		 	
		 	  rgg23                                          & $\approx$8.4M     & $\approx$63.5M \\
		 	  rgg24                                          & $\approx$16.7M    & $\approx$132.6M\\
		 	
		 	  deu                                            & $\approx$4.4M     & $\approx$5.5M \\
		 	  eur                                            & $\approx$18.0M    & $\approx$22.2M \\
		 	
		 	  af\_shell9                                     & $\approx$504K     & $\approx$8.5M \\
		 	  thermal2                                       & $\approx$1.2M     & $\approx$3.7M   \\
		 	  nlr                                            & $\approx$4.2M     & $\approx$12.5M \\

		 	\hline
	\end{tabular}
\vspace*{-1.75cm}
\end{wraptable}

For $u$, it is slightly trickier, but we do not need to recalculate all its gains from scratch since their only source of change is the edge $e$ that connects $u$ and $v$.
Hence, we respectively subtract and add to $g_b(u)$ the corresponding contribution of $e$ before and after the movement of $v$.
We end up doing the update in time $O\big( |I(v)| + |I(v)| * \overline{|R(u)|}  \big)$, where $\overline{|R(u)|}$ is the average of $|R(u)|, \forall \{v,u\} \in I(v)$.

Observe that the quotient graph refinement never needs to check all the gains of a visited node, rather only its gain for a specific PE.
As a consequence, the delta-gain approach
 computes and keeps many more gains than necessary, which is expensive.
In $k$-way FM and multi-try FM, there is another obstacle:
both escape local optima by allowing movements with negative gain.
When a local optimum escape fails, however, they need to go backwards through a whole sequence of movements.
As a consequence, node movements become more frequent than gain checkups. 
Since delta gains are expensive to update and cheap to read, these local searches end up being inappropriate for the delta-gain technique.
For label propagation, this is not the case since the number of node movements is bounded by the number of gain checks.

\section{Experimental Evaluation}
\label{sec:experiments}
\label{s:exp}

\subparagraph*{Methodology.} 
We performed our implementations using the KaHIP framework (using C++) and compiled them using gcc 8.3 with full optimization turned on (-O3 flag). 
All of our experiments were run on a single core of a  machine with  four sixteen-core Intel Xeon Haswell-EX E7-8867 processors running at $2.5$ GHz, $1$ TB of main memory, and \numprint{32768} KB of L2-Cache.
The machine runs Debian GNU/Linux 10 and Linux kernel version 4.19.67-2.

For experiments based on the two-phase approach for tackling GPMP, we solve GPP using KaHIP~\cite{kaffpa}, since it is among the best sequential partitioners regarding solution quality.
To serve our experimental purposes, we use its solution quality configurations \emph{fast} and \emph{eco}, which are described in~\cite{kaffpa}.
From now on, we respectively refer to them as \emph{K(Fast)} and \emph{K(Eco)}.
KaHIP also contains the \emph{top down} approach to solve OPMP, which we use in our experimental comparisons.
We also run Scotch~\cite{Scotch} configured to only use recursive bipartitioning methods and privilege quality over speed.
Starting with a single domain of PEs containing all processes, Scotch recursively bipartitions the processors of a domain into \emph{sub-domains} of PEs while additional procedures such as FM refinement~\cite{fiduccia1982lth} are applied.
We contacted Christopher Walshaw, who informed us that Jostle~\cite{walshaw2000mpm} is not available anymore. Hence, we can not make comparisons against it.

To keep the evaluation simple, we use the following hierarchy configurations for all the experiments: $D=1:10:100$, $\mathcal{S}=4:16:r$, with $r \in \{1,2,3,\ldots,128\}$. Hence, $k=64 \cdot r$.
Depending on the focus of the experiment, we measure running time and/or $J(\mathcal{C},\mathcal{D}, \Pi)$, as defined in Section~\ref{sec:preliminaries}.
We perform ten repetitions of each algorithm using different random seeds for initialization and calculate the arithmetic average of the computed objective functions and running time.
When further averaging over multiple instances, we use the geometric mean in order to give every instance the same influence on the \textit{final score}. 
Unless explicitly stated, we average all results of each algorithm grouped by $k$.
Given a result of algorithm $A$ for $k_o$~PEs, we express its value $\sigma_A$ (which can be objective or running time) using one or more of the following tools:
(i)~\emph{improvement} over an algorithm~$B$, computed as $\big(\frac{\sigma_B}{\sigma_A}-1\big)*100\%$;
(ii)~\emph{ratio}, computed as $\big(\frac{\sigma_A}{\sigma_{max}}\big)$ with $\sigma_{max}$ being the maximum result for $k_o$ among all competitors including $A$;
(iii)~\emph{relative} value over an algorithm~$B$, computed as $\big(\frac{\sigma_A}{\sigma_{B}}\big)$; 
Lastly, we present performance plots (performance profiles). 
These plots relate the running times of all algorithms to the slowest algorithm on a per-instance basis.
For each algorithm, these ratios are sorted in increasing order. The plots show $\big(\frac{\sigma_\text{A}}{\sigma_\text{slowest}}\big)$ on the y-axis. 
A point close to zero indicates that the algorithm was considerably faster than the slowest algorithm. A value of one therefore indicates that the corresponding algorithm has been among the most time consuming algorithms.

\subparagraph*{Instances.}
Our instances come from various sources to test our algorithm.
We use the largest six graphs from Chris Walshaw's benchmark archive~\cite{soper2004combined}.
Graphs derived from sparse matrices have been taken from the SuiteSparse Matrix Collection~\cite{DBLP:journals/toms/DavisH11}. 
We also use graphs from the 10th DIMACS Implementation Challenge~\cite{benchmarksfornetworksanalysis} website. 
Here, \Id{rggX} is a \emph{random geometric graph} with
$2^{X}$ nodes where nodes represent random points in the unit square and edges
connect nodes whose Euclidean distance is below $0.55 \sqrt{ \ln n / n }$.
The graph \Id{delX} is a Delaunay triangulation of $2^{X}$ random points in the unit square. 
The graphs \Id{af_shell9}, \Id{thermal2},  and \Id{nlr} are from the matrix and the numeric section of the DIMACS benchmark set.
The graphs \Id{europe} and \Id{deu} are large road networks of Europe and Germany taken from~\cite{DSSW09}. 
Basic properties of the graphs under consideration can be found in Table~\ref{tab:test_instances_walshaw}.

\subsection{Algorithm Configuration}

In this section, we present a sequence of experiments to test the performance of our algorithmic components regarding solution quality and running time.
Our general goal consists of individually evaluating the effectiveness and significance of each component.
Our specific goal consists of tuning three different configurations of the algorithm based on different principles:
(i)~a \emph{strong} configuration, mostly concerned with maximizing solution quality;
(ii)~a \emph{fast} configuration, mostly concerned with minimizing running time; and
(iii)~an \emph{eco} configuration, which seeks to balance running time and solution quality.
\begin{table}[b]
\centering
\vspace*{-.5cm}
\caption{Various configurations for the evaluation of different initial mapping algorithms.}
\begin{tabular}{|l|r|r|r|r|r|}
\hline
\multicolumn{1}{|r|}{} & \multicolumn{2}{r|}{GPP construction} & \multicolumn{2}{r|}{OPMP construction} & 
\\ \cline{2-5}
\multirow{-2}{*}{Config.} & Std.Bisec. & Multisec. & Identity           & Top Down & \multirow{-2}{*}{$N^{10}_\mathcal{C}$}   \\ \hline \hline
Bsec          & yes                & no           & if $k$ power of 2 & if $k$ not power of 2 & no  \\ \hline
BsecN           & yes                & no           & if $k$ power of 2 & if $k$ not power of 2 & yes \\ \hline
MsecT          & no                 & yes          & no                 & yes                    & no  \\ \hline
MsecTN           & no                 & yes          & no                 & yes                    &  yes   \\ \hline
MsecI          & no                 & yes          & yes                & no                     & no  \\ \hline
MsecIN           & no                 & yes          & yes                & no                     &   yes  \\ \hline
\end{tabular}
\label{tab:res_init_sol}
\end{table}

Our experimental strategy consists of defining a single \emph{focus} aspect of the algorithm for each experiment.
Then, this specific aspect is tested with different components or setups while other parameters  of the algorithm are kept constant.
Initially we use standard components. Then we use the best component found in an experiment the next section.

We begin by focusing on a representative component of the multilevel scheme: 
(i)~initial mapping;
(ii)~local search{}.
Then, we evaluate algorithmic aspects which only affect running time and memory consumption: the distance matrix representation.
{}
The \emph{standard configuration} consists of the matching--based contraction, all local search methods, explicit storage of distance matrix, and no delta-gains updates.
All experiments in this section ran for the six \emph{tuning graphs} from Table \ref{tab:test_instances_walshaw}.

\subparagraph*{Initial Mapping.}

\begin{figure}[t!]
    \captionsetup[subfigure]{justification=centering}
	\centering
	\begin{subfigure}[t]{0.485\textwidth}
		\centering
		\includegraphics[width=\textwidth]{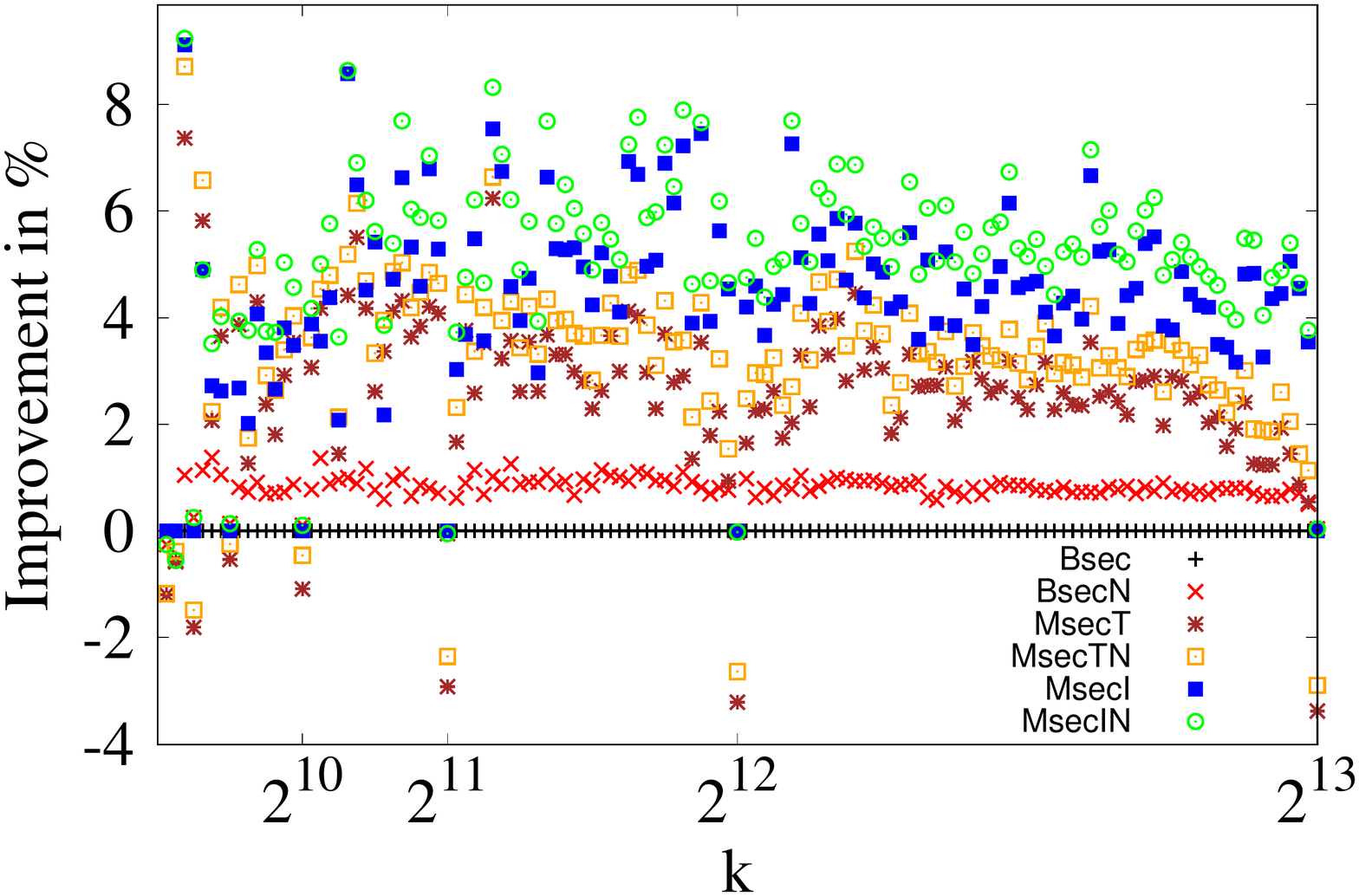}
		\caption{Improvements in objective function over Bsec. Higher is better.}
		\label{fig:InitSolRes}
	\end{subfigure}
	\begin{subfigure}[t]{0.485\textwidth}
		\centering
		\includegraphics[width=\textwidth]{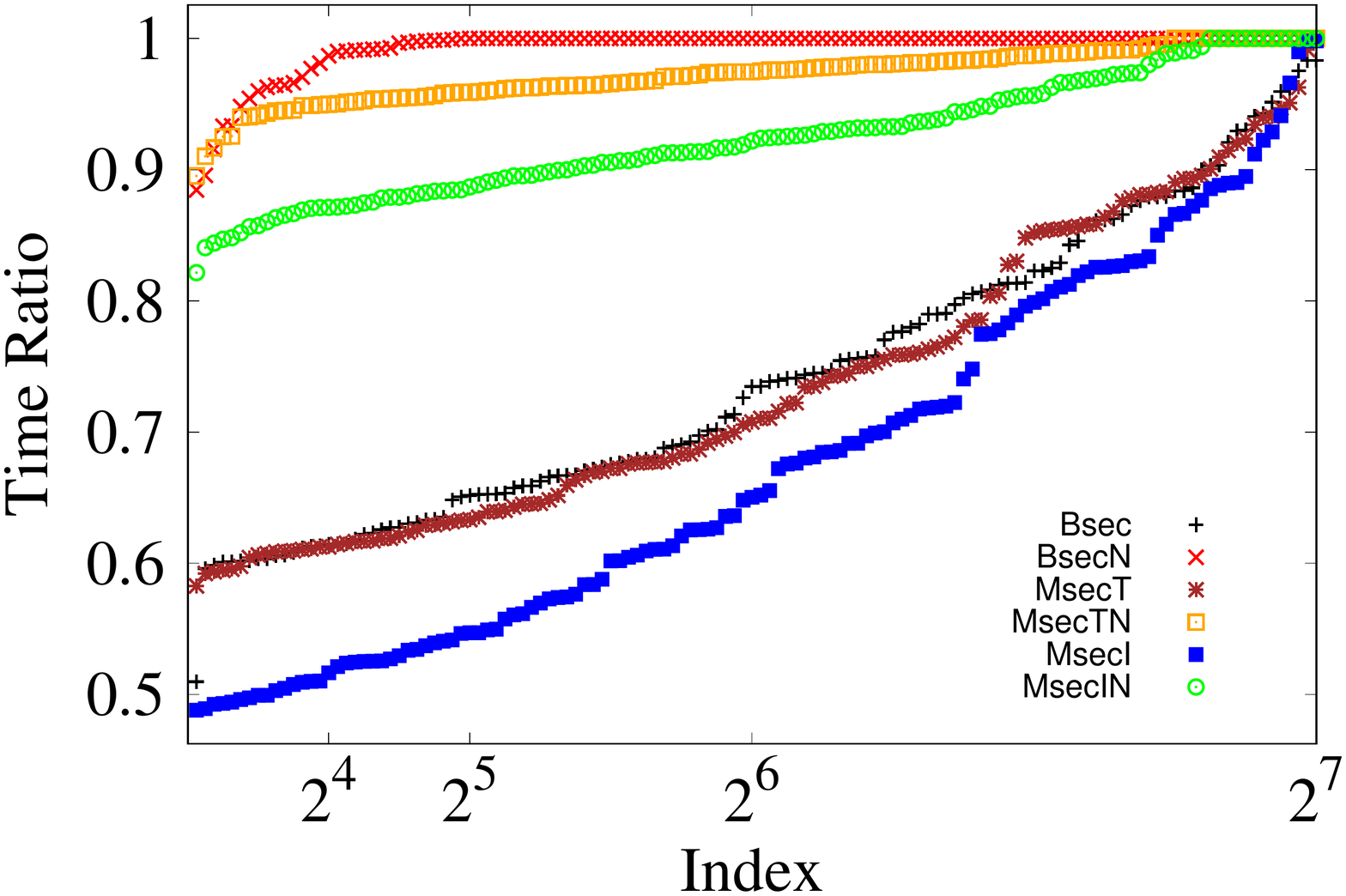}
		\caption{Performance profile for running time (ordered running time ratios). Lower is better.}
		\label{fig:InitSolTim}
	\end{subfigure}

	\caption{Comparing initial mapping algorithms from Table \ref{tab:res_init_sol}.}
	\label{fig:InitSol}
\end{figure}

For the computation of initial mappings, we consider the six configurations listed in Table \ref{tab:res_init_sol}.
Observe that Bsec and BsecN should apply either \emph{identity} or \emph{top down} depending on $k$.
This choice is based on results obtained in \cite{schulz2017better} comparing these two OPMP construction algorithms.
Figure \ref{fig:InitSol} plots the results regarding solution quality and running time for our six configurations.

Looking at solution quality, the configurations using multisection dominate those using standard bisection except for instances having $k$ as a power of 2.
This exception was expected since the standard bisection naturally performs a multisection partition for these instances.
Among the configurations using multisection, \emph{identity} produces overall better solutions than \emph{hierarchy top down}, which is explained by the inherent locality of the multisection approach.
Finally, the $N^{10}_\mathcal{C}$ local search is the least significant factor for solution quality, although it slightly improves solution compared to the similar configurations that skip local search.

In contrast to its low relevance for solution quality, the $N^{10}_\mathcal{C}$ local search is the dominant factor regarding running time.
Observe that the configurations using \emph{identity} are always the fastest ones among those algorithms that either use $N^{10}_\mathcal{C}$ local search or among those that don't.
Hence, the OPMP construction algorithm is the second most relevant factor for running time.
Finally, the partitioning algorithm has little influence over running time, which reflects the rather small time difference between each of the pairs \{BsecN, MsecTN\} and \{Bsec, MsecT\}.

Since MsecIN has the best overall solution quality results, it is the natural choice for \emph{strong}.
Notice that MsecI has the best overall running times, which makes it the perfect choice for \emph{fast}.
Nevertheless, it is also the second best regarding solution quality, which suffices to make it also the best choice for \emph{eco}.

\subparagraph*{Local Search.}

For local search experiments, we start looking at the \emph{fast} algorithm.
To obtain a fast algorithm, we restrict its number of local search methods to one.
Experiments with single local search algorithms do not yield much insight except that  label propagation with delta-gain updates yields a very good trade-off for running time and solution quality.

\begin{figure}
    \captionsetup[subfigure]{justification=centering}
	\centering
	\begin{subfigure}[t]{0.485\textwidth}
		\centering
		\includegraphics[width=\textwidth]{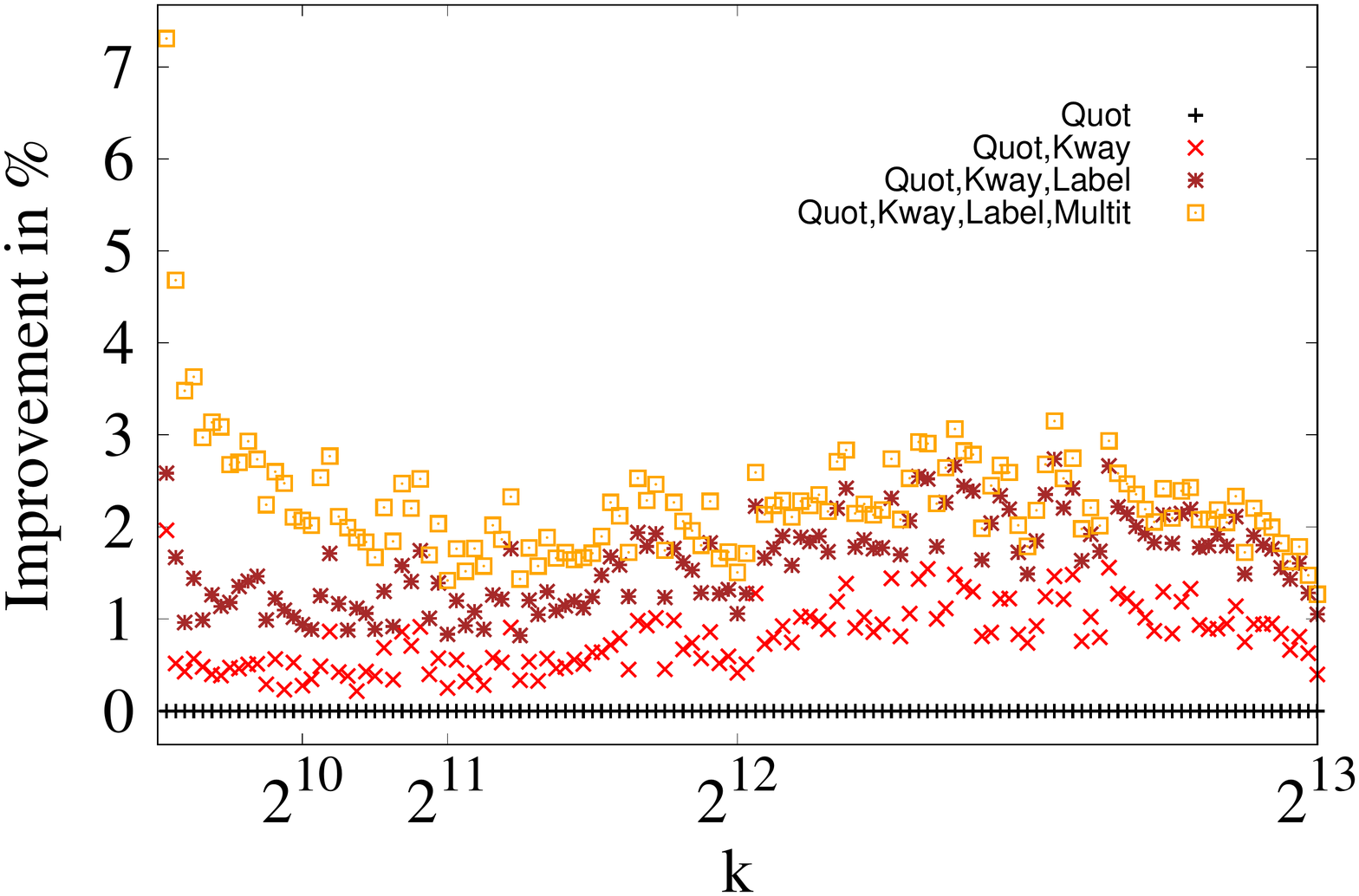}
		\caption{Improvements in objective function over Quot. Higher is better.}
		\label{fig:LSeco_Res}
	\end{subfigure}
	\begin{subfigure}[t]{0.485\textwidth}
		\centering
		\includegraphics[width=\textwidth]{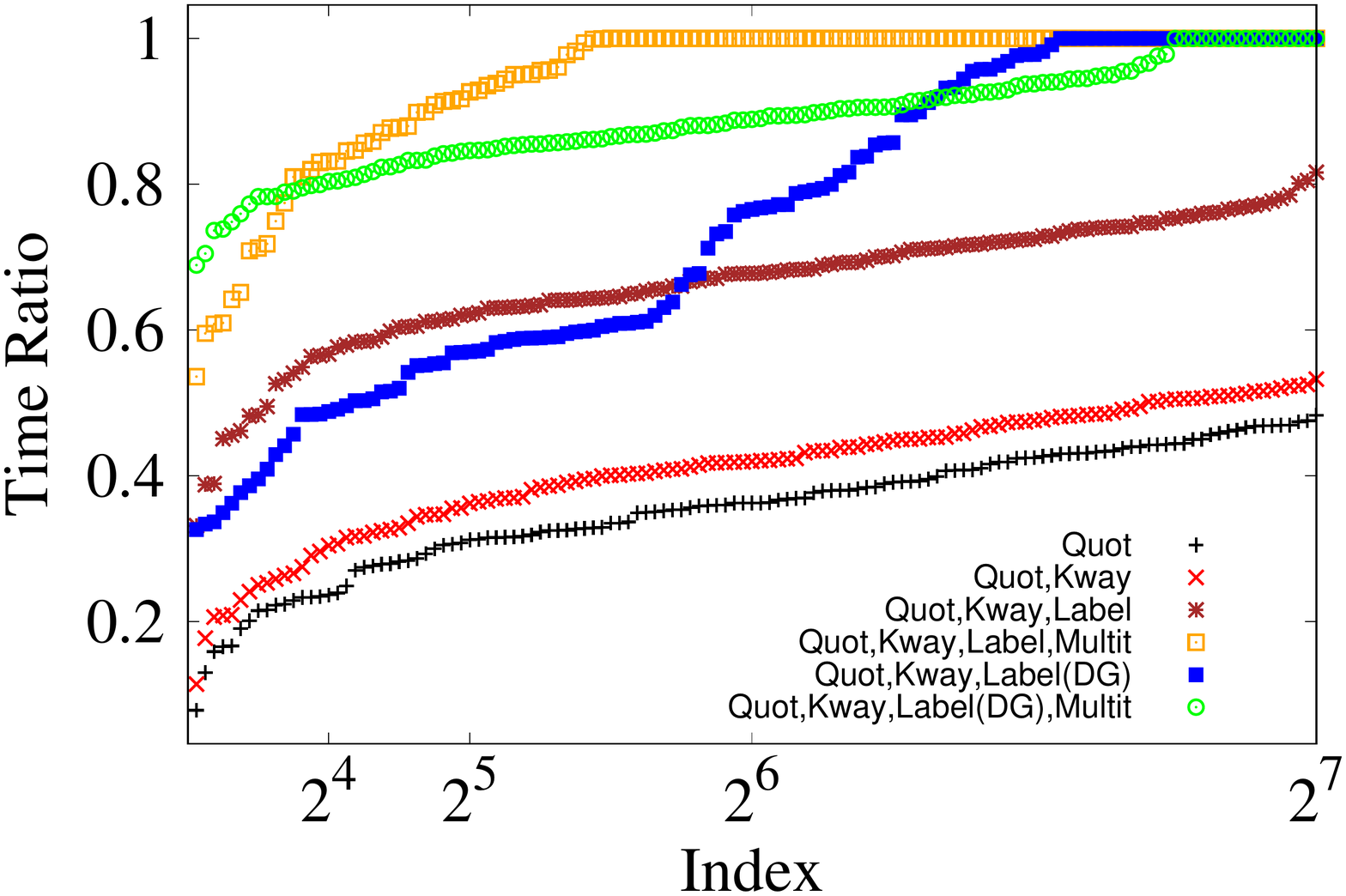}
		\caption{Performance profile for running time (ordered running time ratios). Lower is better.}
		\label{fig:LSeco_Tim}
	\end{subfigure}
	
	\caption{Results for local search experiment of the \emph{eco} algorithm. It comprises four scenarios that represent successive additions of the respective local search methods. 
	In (b), we show two additional scenarios in which delta-gain updates are used (represented by DG).}
	\label{fig:LSeco}
\end{figure}

For the \emph{eco} configuration of our algorithm, we build four configurations by incrementally inserting the local search methods. 
Additionally, we consider two extra configurations equipped with delta-gain updates during label propagation.
Figure \ref{fig:LSeco} summarizes the results concerning these six configurations.
Since the behavior of \emph{strong} in this experiment is equivalent, we omit its results without loss of completeness.

Figure \ref{fig:LSeco} shows that solution quality and running time consistently increase after each consecutive addition of local refinement methods.
Regarding delta gains, running times decrease for some values of $k$ but increase considerably and irregularly for others.
Since this behavior is undesirable for \emph{eco}, we drop delta gains for it.
We also drop delta gains for \emph{strong} since it does not affect solution quality and has negligible influence on running time compared to the $N^{10}_\mathcal{C}$ refinement.
The clear choice for \emph{strong} is the configuration with the four local searches since all of them contribute to incrementally improve solution quality.
For \emph{eco}, we drop only the multi-try FM local search since it adds little to solution quality but significantly increases the running time.

{}

\subparagraph*{Distance Matrix.}

\begin{figure}[t]
    \captionsetup[subfigure]{justification=centering}
	\centering
        \vspace*{-1cm}
		\includegraphics[width=0.5\textwidth]{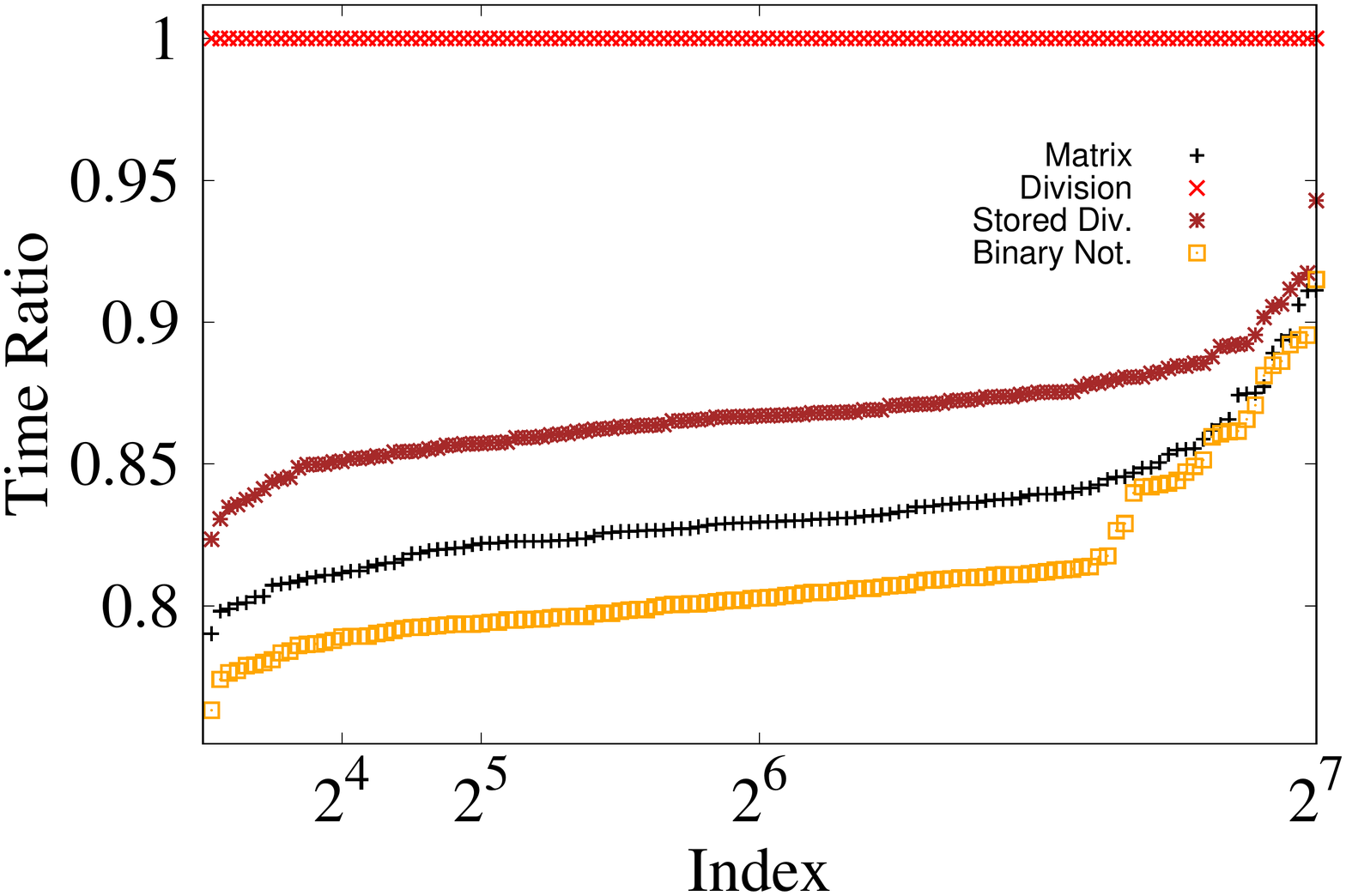}

	\caption{Performance plot for running times of different algorithm configurations (ordered running time ratios for different distance matrix implementations). 
	Four alternative configurations are considered: implicit representations of the distance matrix based on division, stored division, and binary notation, and a reference scenario in which the matrix is explicitly stored. Lower is better.}
	\label{fig:DistMatrix}
\end{figure}

Provided that the objective function is not influenced by the approach used to store or imply the distance matrix, the related experiments only show running time.
We test four configurations: one with each of the three techniques that imply the distance matrix and a reference scenario in which we store the full distance matrix.
Since all configurations of our algorithm display equivalent behavior, we focus on \emph{strong} without loss of generality.
Figure \ref{fig:DistMatrix} plots a running time ratio chart for \emph{strong}.
It is easily understandable that the \emph{binary notation} technique is faster than the \emph{stored division}--based approach, and also that the latter is faster than the \emph{division}--based approach.
On the
   other hand, the \emph{binary notation} outperforms the full-distance matrix approach. While both approaches allow O(1) distance calculations on
   our x86\_64 architecture, accessing the distance matrix incurs a
   memory access.
This leads to frequent cache misses since the $O(k^2)$-sized distance matrix does not fit into the cache of our machine.
Lastly, not using the distance matrix expectedly significantly improves the memory footprint of the algorithm. This especially prounced if the number of blocks gets very large. For example, for $2^{15}$ blocks, not using the distance matrix saves roughly an order of magnitude of necessary main memory.

\subsection{Comparison with State of the Art}

After the tuning step, the three configurations of our algorithm ended up as follows:
(i)~\emph{fast} applies MsecI, label propagation with delta-gain updates, and binary notation;
(ii)~\emph{eco} applies MsecI, quotient graph refinement, $k$-way FM, label propagation, and binary notation; and
(iii)~\emph{strong} applies MsecIN, quotient graph refinement, $k$-way FM, label propagation, multi-try FM, and binary notation.
To improve speed even more, we also include a configuration called \emph{fastest} which applies MsecI as initial mapping, does not use any local search during uncoarsening, and never needs to use information from the distance matrix. 
In this section, we compare them against the best alternative algorithms in the literature.
We report experiments on all graphs listed in Table~\ref{tab:test_instances_walshaw} (excluding the graph used to tune our algorithm). 
In the evaluation we mostly compare the algorithm w.r.t.~the baseline algorithm Müller-Merbach. At the end of this section, we highlight the comparison of various algorithms against each other.

\begin{figure}[t!]
    \captionsetup[subfigure]{justification=centering}
	\centering
	\begin{subfigure}[t]{0.465\textwidth}
		\centering
		\includegraphics[width=\textwidth]{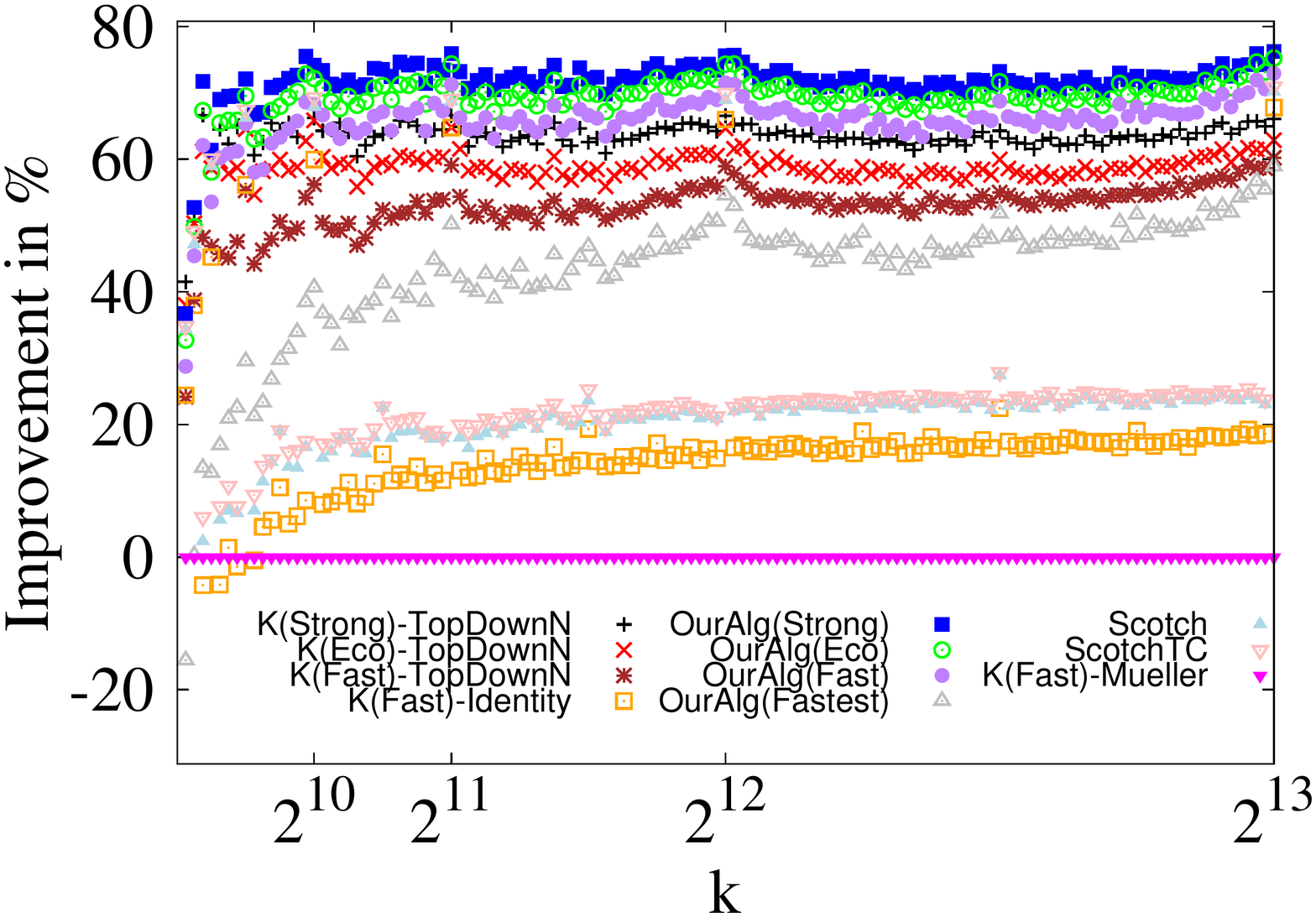}
		\caption{Improvements in objective function over K(Fast)-Müller-Merbach. Higher is better.}
		\label{fig:GenSolRes}
	\end{subfigure}
	\begin{subfigure}[t]{0.505\textwidth}
		\centering
		\includegraphics[width=\textwidth]{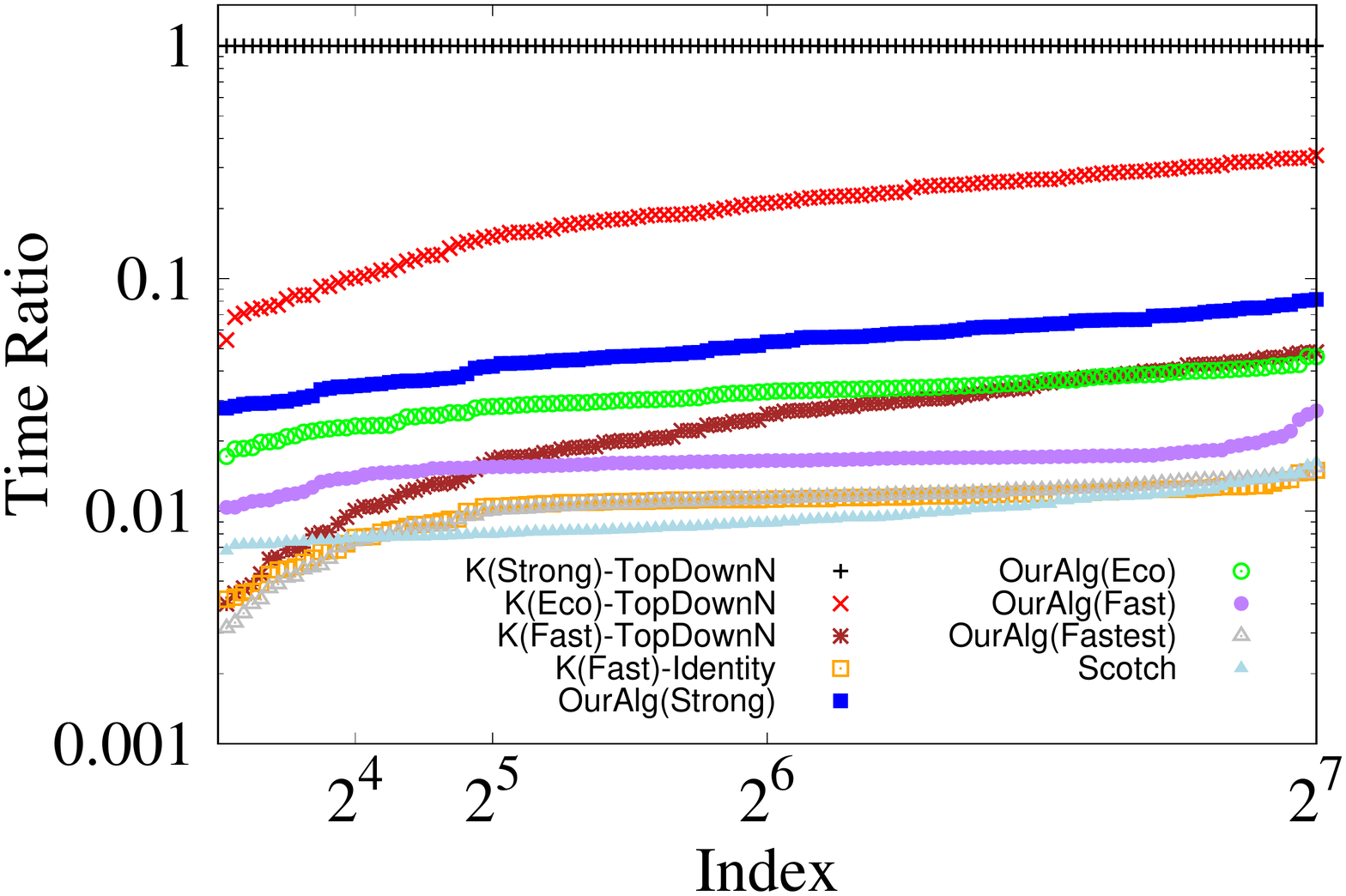}
		\caption{Performance profile for running time. We omit ScotchTC, which equals OurAlg(Strong) in runtime. Lower is better.}
		\label{fig:GenSolTim}
	\end{subfigure}

	\caption{Comparisons against state-of-the-art approaches for GPMP.}
	\label{fig:GenSol}
\end{figure}

We select the most successful algorithms from \cite{schulz2017better} and also Scotch for our comparison:
(i)~\emph{Top down} with $N_{\mathcal{C}}^{d}$ local search (TopDownN), which represent the state-of-the-art for OPMP when $k$ is not a power of 2;
(ii)~\emph{identity} mapping, which (when coupled with the KaHIP multilevel partitioning algorithm) represents the state-of-the-art for GPMP via two-phase approach when $k$ is a power of 2; 
(iii)~the algorithm of \emph{Müller-Merbach}~\cite{muller2013optimale} (Müller-Merbach), whose results are also used as a reference algorithm to calculate solution improvements in~\cite{schulz2017better}; and
(iv)~\emph{Scotch}~\cite{Scotch}.
We run the two-phase appraoches TopDownN, Identity, and Müller-Merbach coupled with K(Fast) as a partitioning algorithm.
Since K(Fast)-TopDownN is our strongest competitor, we additionally couple TopDownN with K(Eco) and K(Strong) to obtain comparable running times.
Recall that these algorithms are non-integrated: they use different quality configurations of KaHIP to partition the graph, compute the coarser communication model and then use TopDownN to compute a one-to-one mapping of blocks to processors (and hence of nodes to processors overall).
Scotch is among the algorithms with best running times in our experiments. 
Hence, we add an algorithm (ScotchTC) which reports the best solution out of multiple runs of Scotch with different random seeds when given the same amount of time to compute a solution as our \emph{strong} configuration has used.
Figure \ref{fig:GenSol} gives an overview over our results.

Regarding solution quality, our algorithms \emph{strong}, \emph{eco}, and \emph{fast} dominate all the other approaches for most values of $k$.
They respectively achieve average improvements of $72\%$, $69\%$, and $66\%$ over Müller-Merbach.
TopDownN coupled with K(Strong), K(Eco), and K(Fast) generally produce the best solutions among our competitors.
Their average improvements over Müller-Merbach are respectively $63\%$, $59\%$, and $53\%$.
Our \emph{fastest} algorithm comes next with an average improvement of $43\%$.
Following are  ScotchTC, Scotch, and K(Fast)-Identity, with improvements $24\%$, $23\%$, and $16\%$, respectively.
When $k$ is a power of 2, our algorithms \emph{strong} and \emph{eco} are also the best, with average improvements over Müller-Merbach of $70\%$, $67\%$, respectively.
They are followed by ScotchTC ($65\%$), Scotch and \emph{fast} ($64\%$ each), K(Strong)-TopDownN ($63\%$), K(Eco)-TopDownN ($62\%$), K(Fast)-Identity ($57\%$), K(Eco)-TopDownN ($53\%$), and our algorithm \emph{fastest} ($40\%$).
Observe that some of our competitors produce solutions with higher average quality when $k$ is a power of 2.
This happens because KaHIP and Scotch are based on recursive bisection schemes, which automatically partition throughout the hierarchy for such values of $k$.

Scotch has the lowest average running time, directly followed by our algorithm \emph{fastest}, K(Fast)-Identity, and our algorithm \emph{fast} (respectively $9\%$, $10\%$, and $73\%$ slower than Scotch on average).
Next, the average running time of K(Fast)-TopDownN is a factor $2.3$ higher than Scotch.
For our algorithms \emph{eco} and \emph{strong}, this factor is respectively $3.3$ and $5.4$.
By definition ScotchTC is also a factor $5.4$ higher than Scotch. 
K(Eco)-TopDownN has a much higher running time ($20.4$ times slower than Scotch).
K(Strong)-TopDownN is our strongest competitor regarding solution quality but the slowest one ($107$ times slower than Scotch).
Our \emph{fast}, \emph{eco}, and \emph{strong} are respectively $62$, $32$, and $20$ times faster than K(Strong)-TopDownN, but still better than it regarding solution quality.

We now highlight various the comparison of various configurations/algorithms.
The previously best approach in terms of overall mapping quality has been K(Strong)-TopDownN.
Our \emph{strong} configuration improves solution quality over K(Strong)-TopDownN by $5.1$\% while being a factor $20$ on average faster.
Our \emph{eco} configuration has roughly $3.6\%$ better quality than K(Strong)-TopDown but is a factor $32$ faster on average. 
Our \emph{fast} configuration still yields $1.3\%$ better solutions on average, and is a factor $62$ faster.
Improved solution quality comes from the fact that the new algorithms are integrated and not two-phase, \ie the multilevel algorithm directly optimize the correct objective. Improvements in running time are achieved since the KaHIP itself (which is used for partitioning) uses even more expensive local search algorithms such as flow-based improvement algorithms or global search schemes like V-cycles. 
Lastly, our \emph{fastest} algorithm is on average $9\%$ slower than Scotch but also improves solution quality over Scotch by $16\%$.

\section{Conclusion}

As high-performance computing systems expand their processing power, there is also a growth regarding number of components, level of parallelism, and sophistication of the topology.
In this work, we tackled the general process mapping problem, which consists of assigning a set of processes to a set of processing elements respecting an imbalance constraint in order to minimize the total communication cost between cores.
Assuming a hierarchically organized topology and a sparse communication matrix containing much more processes than processing elements, we proposed, implemented, tuned, and tested integrated process mapping algorithms to tackle this problem.

We engineered all components of our novel algorithms within a multilevel scheme.
Important ingredients of our algorithms include:
(i) a recursive construction of initial solutions based on multisections throughout the hierarchy of processing elements;
(ii) contraction-uncontraction schemes based on matchings;
(iii) high-quality refinement methods such as label propagation, quotient graph refinement, and very localized local searches;
(iv) a compressed structure to efficiently compute processor distances without storing a distance matrix; and
(v) a memory scheme to keep delta-gain updates during label propagation in order to avoid recomputations of the objective function.

Experimental results indicate that our algorithms are the new state-of-the-art for general process mapping regarding solution quality.
In particular, our algorithms generate much better overall solutions in comparison to any of their competitors while being faster than the previous best algorithm in terms of quality.
Moreover, the best competitor regarding overall solution quality is, at the same time, slower and less effective than the simplest configuration of our algorithm.
Our improvements are mostly due to the integrated multilevel approach combined with high-quality local search algorithms and initial mapping algorithms that split the initial network along the specified system hierarchy.

Important future work includes parallelization as well as the integration of global search schemes and different types of coarsening to improve solution quality further. 
Moreover, we want to investigate the impact that this new technology has on the real performance of applications such as sparse matrix vector multiplications.
Lastly, we plan to release the proposed algorithms in the VieM (\url{http://viem.taa.univie.ac.at/}) and KaHIP (\url{http://algo2.iti.kit.edu/kahip/}) frameworks.

\bibliographystyle{myplainnat}
\bibliography{phdthesiscs}

\end{document}